\documentclass[3p]{elsarticle}
\usepackage[cmex10]{mathtools}
\usepackage{amsfonts}
\usepackage{algorithm}
\usepackage{algorithmic}
\usepackage{url,color}
\newtheorem{theorem}{Theorem}
\newtheorem{conjecture}{Conjecture}
\newtheorem{lemma}{Lemma}
\usepackage{setspace,caption,url}
\newenvironment{myproof}{{\bf Proof.}}{\hfill$\square$}

\newcommand{\st}[0]{\text{s.t.}}
\newcommand{\trace}[1]{\operatorname{Trace}\left(#1\right)}
\newcommand{\rank}[1]{\operatorname{Rank}\left(#1\right)}
\newcommand{\diag}[1]{\operatorname{Diag}\left(#1\right)}
\newcommand{\sign}[1]{\operatorname{Sign}\left(#1\right)}
\newcommand{\expect}[1]{\mathbb{E}\left(#1\right)}
\newcommand{\R}[0]{\mathbb{R}}
\newcommand{\C}[0]{\mathbb{C}}
\newcommand{\Pb}[1]{\mathbb{P}\left\{#1\right\}}
\newcommand{\I}[0]{\mathcal{I}}
\newcommand{\cF}[0]{\mathcal{F}}

\newcommand{\vs}[0]{\mathbf{s}}
\newcommand{\vt}[0]{\mathbf{t}}
\newcommand{\ve}[0]{\mathbf{e}}
\newcommand{\vw}[0]{\mathbf{w}}
\newcommand{\vx}[0]{\mathbf{x}}
\newcommand{\ma}[0]{\mathbf{A}}
\newcommand{\mb}[0]{\mathbf{B}}
\newcommand{\mc}[0]{\mathbf{C}}
\newcommand{\me}[0]{\mathbf{E}}
\newcommand{\mf}[0]{\mathbf{F}}
\newcommand{\ms}[0]{\mathbf{S}}

\newcommand{\hs}[0]{\hat{\vs}}
\newcommand{\ts}[0]{\tilde{\vs}}

\newcommand{\hS}[0]{\widehat{\ms}}

\begin{document}
\begin{frontmatter}

\title{Design of Spectrally Shaped Binary Sequences\\ via Randomized Convex Relaxation}
\tnotetext[titleNote]{This work is partially supported by the National Science Foundation under
grants ECCS-1201835 and AST-1547278. Portions of this work were presented at the Asilomar Conference
on Signals, Systems and Computation~\cite{Mo2015Design-of-Spect}.}

\author[mainAddress]{Dian Mo}
\ead{mo@umass.edu}
\author[mainAddress]{Marco F. Duarte}
\ead{mduarte@ecs.umass.edu}
\address[mainAddress]{Department of Electrical and Computer Engineering, University of
Massachusetts, Amherst, MA 01003}

\begin{abstract}
    Wideband communication receivers often deal with the problems of detecting weak signals from
    distant sources received together with strong nearby interferers. When the techniques of random
    modulation are used in communication system receivers, one can design a spectrally shaped
    sequence that mitigates interferer bands while preserving message bands. Common implementation
    constraints require sequence quantization, which turns the design problem formulation to an
    integer optimization problem solved using a semidefinite program on a matrix that is restricted
    to have rank one. Common approximation schemes for this problem are not amenable due to the
    distortion to the spectrum caused by the required quantization. We propose a method that
    leverages a randomized projection and quantization of the solution of a semidefinite program, an
    approach that has been previously used for related integer programs. We provide a theoretical
    and numerical analysis on the feasibility and quality of the approximation provided by the
    proposed approach. Furthermore, numerical simulations show that our proposed approach returns
    the same sequence as an exhaustive search (when feasible), showcasing its accuracy and
    efficiency. Furthermore, our proposed method succeeds in finding suitable spectrally shaped
    sequences for cases where exhaustive search is not feasible, achieving better performance than
    existing alternatives.
\end{abstract}
\begin{keyword}
    wideband communication, sequence design, integer programming, semidefinite programming
    relaxation, rank-one approximation, randomized projection
\end{keyword}

\end{frontmatter}

\section{Introduction}

Receivers for emerging wireless communication systems are expected to deal with a very wide spectrum
and adaptively choose which parts of it to extract. The intense demand on the available spectrum for
commercial users force many devices to share the spectrum~\cite{Salzman2001Proceedings-of-,
rowe2014spectrally}. As a result, wideband communication systems have caused interference for
applications using overlapping regions~\cite{Aubry2014Radar-waveform-, Aubry2015A-new-radar-wav}.
Thus, a major issue for wideband communication receivers is to process spectra having very weak
signals from a distant source mixed with strong signals from nearby sources. Practical
nonlinearities in receiver circuits make the separation of such signals a key barrier for wideband
communication systems since unfiltered interferers are large enough to cause distortion that can
mask weaker signals.

Recently, random sequences for wideband signal modulation have been employed in the realization of
communication system receivers~\cite{laska2007theory, Mishali2009Blind-Multiband, tropp2010beyond}.
In essence, the signal after random modulation contains a baseband spectrum that is the linear
combination of all frequency components of the input signal. Thus, using a large enough number of
modulation branches allows for successful recovery of the wideband signal, where the number of
necessary branches is determined by the occupancy of the spectrum. The resulting multi-branch
modulation system can be abstracted as an all-pass filter that preserves all frequency components of
interest from the input signal.

In many cases, when the locations of one or more interferers is known, the modulation to null out
the interferer band is desirable to reduce the distortion due to nonlinearities. Therefore, it is
promising to replace the pseudo-random sequence with a spectrally shaped sequence that effectively
implements a notch filter to suppress interferers. In addition to the strong interferer case
described earlier, a similar problem arises in dynamic spectrum management (DSM), an approach that
allows for flexibility in spectrum use. DSM attempts to determine the frequencies being used by
previous or licensed applications and selects an optimal subset from the remaining frequencies for
new or unlicensed applications. The signals for unlicensed applications should be optimized to
minimize the interference with licensed signals while keeping their own capacities. More
specifically, for Direct Sequence Spread Spectrum, a designed spreading code with particular
spectral characteristic has been used to shape the unlicensed power
spectra~\cite{clancy2006spectrum}. Another similar problem arises in active sensing, which obtains
valuable information of targets or the propagation medium by sending probing waveforms toward an
area of interest~\cite{Li2009MIMO-Radar-Sign, Zhao2013Enhanced-multis, Liang2014On-Designing-th}. A
well-designed waveform is crucial to the performance of active sensing.

To maximize the power efficiency in modulation, it is a standard requirement that the modulating
waveform is constrained to be unimodular. Additionally, it is desirable that the waveform possess
some specific spectrum magnitude. Such a choice improves the target detection performance in active
sensing~\cite{He2012Waveform-Design}. A common performance criterion gives preference to low
autocorrelation sidelobes in the time domain or a flat spectrum in the frequency domain, since the
autocorrelation function and the power spectral density form a Fourier transform pair. Metrics of low 
autocorrelation previously used in sequence design include the integrated sidelobe level (ISL) and the 
peak sidelobe level (PSL).  Based on this
relationship and these metrics, an iterative method has been developed to design unimodular waveforms with flat
spectrum and impulse-like autocorrelation function by alternatively determining the waveform and an
auxiliary phase~\cite{Stoica2009New-Algorithms-, Stoica2009On-Designing-Se}.
More recently, additional methods including majorization-minimization, coordinate descent, and alternating direction method of 
multipliers have been used in the literature to optimize the values of the ISL, PSL, and their weighted 
variants~\cite{Song2015Optimization-Me, Song2016WISL,Kerahroodi2017PSL, Zhao2017Autocorr, 
Li2018Unimodular, Liang2016Unimodular-Sequ}. 

Due to the fact that many devices are
forced to coexist, the waveforms are constrained to have spectral nulls in specific bands to keep
the mutual interference within acceptable levels. Based on the iterative method
of~\cite{Stoica2009New-Algorithms-, Stoica2009On-Designing-Se}, the SHAPE algorithm designs
sequences that simultaneously approach the desired spectrum magnitude and satisfy the envelope
constraint~\cite{Liang2014On-Designing-th, rowe2014spectrally}. A sequence is obtained by
alternatively searching the frequency and time domains to minimize the estimation error while
meeting the two aforementioned criteria. In~\cite{Liang2015Waveform-Design}, a Lagrange programming
neural network (LPNN) algorithm~\cite{Zhang1992Lagrange-progra}, based on nonlinear constrained
optimization, is applied to design waveforms with unit modulus and spectral constrains. While it is possible to
modify some of the algorithms in the literature to switch from an unimodular constraint to a binary
constraint for the sequences, our numerical simulations in the sequel show that such changes result
in significant performance losses in modulation.

In this work, we consider the problem of designing a binary sequence for modulation that is tailored
for message preservation and interferer cancellation in order to allow for a simple system
implementation~\cite{Mo2015Design-of-Spect}. More specifically, we aim to find binary sequences with
sufficiently large spectrum magnitudes for the {\em message band\/} (i.e., for the frequencies where
the message lies) while keeping sufficiently small magnitudes for the {\em interferer band\/} (i.e.,
frequencies where the interferer may exist).  This design of spectrally shaped binary sequences can
be formulated as a quadratically-constrained quadratic program (QCQP) with an objective function 
that measures message preservation and both equality constraints
(for binary quantization) and inequality constraints (for interferer mitigation). This problem is
NP-hard~\cite{luo2007approximation} and an exhaustive search for the optimal solution is
computationally prohibitive beyond very small sequence lengths. It is well known that such an
optimization problem can be written as a convex semidefinite program (SDP) with a non-convex
rank-one constraint~\cite{Shor1987Quadratic-optim, Lovasz1991Cones-of-Matric}. Relaxing the SDP by
dropping the rank constraint makes the problem solvable but also requires a procedure to reduce the
SDP solution to a rank-one matrix in the highly likely case that the rank constraint is not met.

Goemans and Williamson~\cite{goemans1995improved} proposed a randomized projection and binary
quantization method to provide improved approximations for the Maximum Cut problem, which is another
important binary optimization problem involving a QCQP\@; however, the Maximum Cut problem features
only equality constraints. The randomized projection returns a rank-one approximate solution that
optimizes the value of the quadratic objective function in expectation. A best approximation is
selected from repeated instances of the randomized projection by using a selection criterion. There
exists a significant literature that has extended this method to approximately solve many similar
optimization problems~\cite{Fu1998Approximation-A, Nesterov1998Semidefinite-re,Ye1999ApproximatingG,
ye1999approximating, Nemirovski1999On-maximization, zhang2000quadratic, luo2007approximation}. For
example, the randomized projection is proven to approximately solve the QCQP with only inequality
constraints~\cite{Ye1999ApproximatingG, Nemirovski1999On-maximization, luo2007approximation} as well
as the QCQP with both inequality constraints and equality constraints meeting a special diagonal
structure~\cite{ye1999approximating, zhang2000quadratic}. In the field of radar waveform design, the
introduction of SDP relaxation and randomized projections has already shown that accurate and
sometimes near-optimal approximations of complex-valued sequences are 
feasible~\cite{Maio2008Code-Design-to-, Maio2009Design-of-Phase, Maio2011Design-of-Optim, 
Cui2014MIMO-Radar-Wave, Aubry2016Forcing-Multipl}. However, the gap between the complex-valued 
sequence design in radar and the binary sequence design considered here prevents these analyses in
the literature from applying on the problems we consider here.

In this paper, we present an algorithm to design binary sequences targeted to meet a specific
spectrum shape. The algorithm is based on an SDP relaxation of a QCQP followed by a randomized
projection and binary quantization, an approach that is inspired by~\cite{goemans1995improved}. Our main 
contributions can be detailed as follows. First, we propose a spectrally shaped binary sequence design approach 
based on optimization via a QCQP\@. Second, we extend the randomized projection and binary quantization method 
of~\cite{goemans1995improved} to our QCQP, which features both equality and inequality constraints. Third, 
we provide analytical and numerical results that study the feasibility of the sequences obtained from the proposed 
randomization, as well as the quality of the approximation achieved by the proposed algorithm. Fourth, we propose 
several custom score functions for the sequences obtained from randomization that allow for an improved 
selection of binary sequences that achieve both message preservation and interference rejection. Finally, we
present numerical simulations that perform a comparison between an exhaustive search and the
proposed sequence design method when the sizes that are sufficiently small to make exhaustive search
feasible. The numerical results verify that our proposed method finds the optimal binary sequences.
We also provide numerical simulations that show the advantages of the proposed algorithm against
algorithms from the literature that have been modified when necessary to provide binary sequence
designs.

This paper is organized as follows. In Section~\ref{section:background}, we give a brief
introduction to the SHAPE and LPNN algorithms for unimodular sequence design and present simple
changes to both algorithms in order to make them suitable for binary sequence design. In
Section~\ref{section:QCQP}, we provide a brief summary of QCQPs and existing approaches to solving
QCQPs via SDP relaxation and randomized projection. In Section~\ref{section:design}, we present and
analyze the use of optimization and randomized projection to design spectrally shaped binary
sequences; furthermore, we provide alternative criterion for sequence selection so that the
resulting sequence allows for better interferer rejection. In Section~\ref{section:results}, we
present numerical simulations to validate the analysis in Section~\ref{section:design}. Finally, we
provide a discussion and conclusions in Section~\ref{section:summary}.

\section{Background}\label{section:background}

To the best of our knowledge, there is no method in the literature that directly addresses the
binary sequence design problem. Nonetheless, we summarize in this section two existing approaches
for the problem of sequence design with unit modulus. We include those two approaches since they can 
be easily modified to design binary sequences by changing the optimization constraints. For other 
approaches to unimodular sequence design, such a change is not straightforward~\cite{Maio2008Code-Design-to-,
Maio2009Design-of-Phase, Maio2011Design-of-Optim, Cui2014MIMO-Radar-Wave, Aubry2016Forcing-Multipl}.
We will use the original and proposed modified algorithms for sequence design in the numerical
simulations presented in Section~\ref{section:results}.

\subsection{SHAPE Algorithm}
The SHAPE algorithm aims to find an unimodular sequence $s$ whose spectrum $vx$ has magnitude that
meets both an upper bound $f _i$ and a lower bound $g _i$ ($i = 1, 2, \dots, N$),
repectively~\cite{Liang2014On-Designing-th, rowe2014spectrally}:
\begin{align}
    \hs = \arg \min _{\vs, \vx \in \C ^N, \alpha \in \C} &\quad \|\cF ^H \vs - \alpha \vx\| _2 ^2 \notag \\
    \st & \quad |s _i| ^2 = 1, i = 1, 2, \dots, N \notag \\
        & \quad |x _i|  \le f _i, i = 1, 2, \dots, N \notag \\
        & \quad |x _i|  \ge g _i, i = 1, 2, \dots, N,
    \label{equation:SHAPE}
\end{align}
where $\cF$ collects all elements of a discrete Fourier transform basis and $\alpha$ is a scalar
factor accounting for the possible energy mismatch between the sequence and the constraints. The
SHAPE algorithm solves (\ref{equation:SHAPE}) using an iterative approach with the following three
main steps.

\begin{enumerate}
    \item Given $\vs$ and $\alpha$, find the spectrum $\vx$:
        \begin{align}
            \hat{\vx} = \arg \min _{\vx \in \C ^N} &\quad \|\cF ^H \vs - \alpha \vx\| _2 ^2 \notag \\
            \st & \quad |x _i|  \le f _i, i = 1, 2, \dots, N \notag \\
                & \quad |x _i| \ge g _i, i = 1, 2, \dots, N.
        \end{align}
The solution is given by
\begin{align}
\hat{x} _i = \left\{ \begin{aligned}
        &f _i \frac{\cF _i ^H \vs / \alpha}{\left| \cF _i ^H \vs /\alpha \right|}, & \left| \cF _i
        ^H \vs /\alpha \right| > f _i \\
        &g _i \frac{\cF _i ^H \vs / \alpha}{\left| \cF _i ^H \vs /\alpha \right|}, & \left| \cF _i
        ^H \vs /\alpha \right| < g _i \\
        &\cF _i ^H \vs / \alpha, &\text{otherwise}
    \end{aligned} \right.,
\end{align}
$i = 1, 2, \dots, N$, where $\cF _i$ denotes the $i ^{\text{th}}$ column of $\cF$.

\item Given $\vs$ and $\vx$, find the factor $\alpha$:
    \begin{align}
        \hat{\alpha} = \arg \min _{\alpha \in \C} \|\cF ^H \vs - \alpha \vx\| _2 ^2.
    \end{align}
The solution is given by
\begin{align}
    \hat{\alpha} = \frac{\vx ^H \cF ^H \vs}{\left\| \vx \right\| _2 ^2}.
\end{align}
\item Given $\alpha$ and $\vx$, find the sequence $\vs$:
    \begin{align}
        \hs =  \arg \min _{\vs \in \C ^N} &\quad \left\| \cF ^H \vs - \alpha \vx \right\| _2 ^2 \notag \\
        \st &\quad \left| s _i \right| ^2 = 1, i = 1, 2, \dots, N.
        \label{eq:shape-uni}
    \end{align}
The solution is given by
\begin{align}
    \hs _i = \frac{\alpha \cF _i ^H \vx}{\left| \alpha \cF _i ^H \vx \right|}, i = 1, 2, \dots, N.
\end{align}
\end{enumerate}

A straightforward change to the SHAPE algorithm for binary sequence design is to force the desired
sequence $\vs$ to be real in (\ref{equation:SHAPE}). This change is equivalent to replacing the
optimization (\ref{eq:shape-uni}) with the binary constraint problem
\begin{align}
    \hs =  \arg \min _{\vs \in \R ^N} &\quad \left\| \cF ^H \vs - \alpha \vx \right\| _2 ^2 \notag \\
    \st &\quad \left| s _i \right| ^2 = 1, i = 1, 2, \dots, N.
\end{align}
In other words, the resulting binary sequence can be obtained as
\begin{align}
    \hs _i = \sign{\mathfrak{R}(\alpha \cF _i ^H \vx)}, i = 1, 2, \dots, N,
\end{align}
where $\mathfrak{R}(\cdot)$ denotes the real part of a complex number and $\sign{\cdot}$ denotes the
sign of a real number.

\subsection{LPNN Algorithm}
In~\cite{Liang2015Waveform-Design}, a Lagrange programming neural network (LPNN) for unimodular
sequence design with target spectrum $\vx$ is formulated as follows:
\begin{align}
    \hs = \arg \min _{\vs \in \C ^N, \alpha \in \R}
    &\quad \sum _ {i = 1} ^N w _i {\left( \left| \cF _i ^H \vs \right| ^2 - \alpha x _i \right)} ^2
    +  c _0 \sum _{i = 1} ^N {\left( \left| \ve _i ^T \vs \right|^2 - 1\right)} ^2 \notag \\
    \st 
    &\quad \left| \ve _ i ^T \vs \right| ^2 = 1, i = 1, 2, \dots, N.
    \label{equation:LPNN}
\end{align}
Here $\ve _i$ denotes the canonical column vector whose $i ^{\text{th}}$ entry is $1$ and others are
$0$, and $w _i$ are weights for each frequency component. The second term in the objective function
is the augmented term to improve the algorithm's convexity and stability. By separating the real and
imaginary parts of the matrices and vectors in the equation as
\begin{align*}
\vt = \begin{bmatrix} \mathfrak{R}\{\vs\} \\ \mathfrak{I}\{\vs\} \end{bmatrix}, \quad
\mf _i = \begin{bmatrix} \mathfrak{R}\{\cF _i\} & \mathfrak{I}\{\cF _i\} \\ -\mathfrak{I}\{\cF _i\} &\mathfrak{R}\{\cF _i\} \end{bmatrix}, \quad
    \me _i = \begin{bmatrix} \ve_i & \mathbf{0} \\ \mathbf{0} & \ve _i \end{bmatrix},
\end{align*}
the complex-valued optimization (\ref{equation:LPNN}) is transformed into the real-valued
optimization
\begin{align}
    \min _{\vt \in \R ^{2 N}, \alpha \in \R}
    & \quad \sum _{i = 1} ^N w _i {\left( \vt ^T \mf _i \mf _i ^T \vt - \alpha x _i \right)} ^2 + \quad c_0 \sum _{i = 1} ^N {\left( \vt ^T \me _i \me _i ^T \vt - 1 \right)} ^2 \notag \\
    \st & \quad \vt ^T \me _i \me _i ^T \vt = 1, i = 1, 2, \dots, N.
\end{align}
The Lagrangian function for this problem is set up as 
\begin{align}
    \mathcal{L} \left( \vt, \alpha, \boldsymbol{\mu} \right) 
    &= \sum _{i = 1} ^N w _i {\left( \vt ^T \mf _i \mf _i ^T \vt - \alpha x _i \right)} ^2 + c_0 \sum _{i = 1} ^N {\left( \vt ^T \me _i \me _i ^T \vt - 1 \right)} ^2 + \sum _{i = 1} ^N \mu _i \left( \vt ^T \me _i \me _i ^T \vt - 1 \right),
\end{align}
where $\mu _i$ is the Lagrange multiplier. The LPNN then computes increments for the parameters and
solution of this problem as follows:
\begin{align}
    \Delta \vt 
    &= - \frac{\partial \mathcal{L}}{\partial \vt} = - 4 \sum _{i = 1} ^N w _i \left( \vt ^T \mf _i \mf _i ^T \bar{s} - \alpha x _i \right) \mf _i
    \mf _i ^T  - 4 c _0 \sum _{i = 1} ^N \left( \vt ^T \me _i \me _i ^T \vt - 1 \right) \me _i \me _i
    ^T  - 2 \sum _{i = 1} ^N \mu _i \me _i \me _i ^T \vt,
    \label{equation:dynamic1}
\end{align}
\begin{align}
    \Delta \alpha 
    &= - \frac{\partial \mathcal{L}}{\partial \alpha} = 2 \sum _{i = 1} ^N w _i \left( \vt ^T \mf _i
    \mf _i ^T \vt - \alpha x _i \right) x_i, \\
    \Delta \mu _i 
    &= - \frac{\partial \mathcal{L}}{\partial \mu _i} = \vt ^T \me _i \me _i ^T \vt - 1.
\end{align}
The LPNN algorithm initializes the so-called neurons $\vt$, $\alpha$, $\boldsymbol{\mu}$ randomly. The
neurons are updated using the increments above at each iteration $k$:
\begin{align}
    \vt ^{k + 1} &= \vt ^k + \rho \Delta \vt, \\
    \alpha ^{k + 1} &= \alpha ^k + \rho \Delta \alpha, \\
    \boldsymbol{\mu} ^{k + 1} &= \boldsymbol{\mu} ^k + \rho \Delta \boldsymbol{\mu}.
    \label{equation:update3}
\end{align}
Finally, the unimodular sequence $\hs$ is constructed by taking first and last $N$ entries of
$\bar{s}$ as its real and imaginary parts, respectively.

The LPNN algorithm can be modified to provide binary sequences by changing the $s \in \C^N$
constraint in (\ref{equation:LPNN}) to $s \in \R^N$, making (\ref{equation:LPNN}) a real-valued
optimization. Thus, we can directly obtain the dynamics for the neurons $s$, $\alpha$, and $\mu$ by
replacing $\vt$ to $\vs$, $\mf _i$ and $\mf _i ^ T$ to $\cF _i$ and $\cF _i ^H$, and $\me _i$ to $\ve _i$ in
(\ref{equation:dynamic1}-\ref{equation:update3}).

\section{Quadratically Constrained Quadratic Programming}\label{section:QCQP}

In this section, we  summarize approaches to approximately solve QCQPs and provide available
analytical frameworks for the approximation performance. The approaches described in this section
originate from the seminal paper~\cite{goemans1995improved}, with extensions to several related
problems~\cite{Fu1998Approximation-A, Nesterov1998Semidefinite-re,Ye1999ApproximatingG,
ye1999approximating, Nemirovski1999On-maximization, zhang2000quadratic, luo2007approximation}. In
general, a QCQP problem can be written as
\begin{align}
    \hs = \arg \max _{\vs \in \mathbb{R} ^N} \quad 
    f \left( \vs \right) &= \vs ^T \ma \vs \notag \\
    \st \quad
    g _k \left( \vs \right) &= \vs ^T \mb _k \vs \le \alpha _k, \quad k \in \mathcal{I}, \notag \\
    h _k \left( \vs \right) &= \vs ^T \mc _k \vs = \beta _k, \quad k \in \mathcal{E},
    \label{equation:QCQP}
\end{align}
where $\ma$, $\mb _k$ and $\mc _k$ are characteristic matrices for the objective function $f$, the
inequality constraint function $g _k$ and the equality constraint function $h _k$, respectively.
Specific instances of this QCQP, placing different conditions in the involved matrices, have been
studied in the literature; we focus on the following specific classes.
\renewcommand{\labelenumi}{(\Roman{enumi})}
\begin{enumerate}
    \item There are no inequality constraints (i.e., $|\mathcal{I}| = 0$), $|\mathcal{E}| = N$, and
        $\mc _k = \ve _k \ve _k ^T$. When all $\beta _k = 1$, the equality constraints essentially enforce
        a binary constraint to the solution $s$, which has been used to solve the Maximum Cut
        problem~\cite{goemans1995improved, Maio2009Design-of-Phase, Cui2014MIMO-Radar-Wave}.
    \item There are no equality constraints (i.e., $|\mathcal{E}| = 0$) and $\mb _k$ are all positive
        semidefinite. The feasible set $\{\vs | \vs ^T \mb _k \vs \le \alpha _k, k \in \mathcal{I}\}$ is an
        intersection of ellipsoids with common center~\cite{Ye1999ApproximatingG,
        Nemirovski1999On-maximization, luo2007approximation, Aubry2016Forcing-Multipl}.
    \item There is only one inequality constraint and the characteristic matrix $\mb$ is diagonal,
        $|\mathcal{E}| = N$, and $\mc _k = \ve _k \ve _k ^T$~\cite{ye1999approximating,
        zhang2000quadratic, Maio2011Design-of-Optim}. This class is equivalent to the first class
        when the feasible set is not empty, due to the fact that any solution that satisfies the
        equality constraints (and thus is a vertex of a high-dimensional hypercube) will always
        lie inside the high-dimensional ball described by the inequality constraint.
\end{enumerate}

We will show in the sequel that the proposed binary sequence design can be formed as a QCQP with both
equality constraints and inequality constraints with more general structure, thus not belong to any
of the three classes described here.

\subsection{Semidefinite Programming Relaxation}

Solving a QCQP is NP-hard~\cite{luo2007approximation}. Most optimization methods for QCQP are based
on a relaxation of the problems where an upper bound of of the objective value for the optimal
solution is computed. The SDP relaxation has been an attractive approach due to its potential to
find a good approximate solution for many QCQPs, including the specific classes mentioned above.

By lifting $\vs$ to a symmetric matrix $\ms = \vs \vs ^T \in \mathbb{R} ^{N \times N}$ with
$\rank{\ms} = 1$, the objective function $f$ in (\ref{equation:QCQP}) has a linear representation
with respect to $\ms$. Therefore, the QCQP in (\ref{equation:QCQP}) can be expressed equivalently as
\begin{align}
    \widehat{\ms} = \arg \max _{\ms \in \mathbb{S} ^N} \quad 
    & \trace{\ma \ms} \notag \\
    \st \quad
    & \trace{\mb _k \ms} \le \alpha _k, \quad k \in \mathcal{I}, \notag \\
    & \trace{\mc _k \ms} = \beta _k, \quad k \in \mathcal{E}, \notag \\
    & \rank{\ms} = 1,
    \label{equation:SDPR}
\end{align}
where $\mathbb{S} ^N$ represents the set of all $N$-dimensional positive semidefinite matrices.
Given that such matrices are positive semidefinite and rank-one, any feasible solution $\ms$ to
(\ref{equation:SDPR}) can be factorized as $\vs \vs ^T$ such that $\vs$ is an feasible solution to
(\ref{equation:QCQP}).

Though (\ref{equation:SDPR}) is as difficult to solve as (\ref{equation:QCQP}), it indicates
that the only non-convex constraint is the rank constraint and that the objective function and all
other constraints are convex with respect to $\ms$ when $\ma$, $\mb _k$, and $\mc _k$ are all positive
semidefinite. Thus the SDP relaxation of (\ref{equation:QCQP}) is obtained by dropping the rank
constraint:
\begin{align}
    \widehat{\ms} = \arg \max _{\ms \in \mathbb{S} ^N} \quad 
    f \left( \ms \right) &= \trace{\ma \ms} \notag \\
    \st \quad
    g _k \left( \ms \right) &= \trace{\mb _k \ms} \le \alpha _k, \quad k \in \mathcal{I}, \notag \\
    h _k \left( \ms \right) &= \trace{\mc _k \ms} = \beta _k, \quad k \in \mathcal{E}.
    \label{equation:SDPC}
\end{align}
The resulting convex problem (\ref{equation:SDPC}) can be efficiently solved, e.g., by
interior-point methods~\cite{Potra2000Interior-Point-}.

\subsection{Randomized Projection}

After solving the SDP relaxation, the next important step is to extract a feasible solution $\ts$ to
(\ref{equation:QCQP}) from the optimal solution $\hS$ resulting from (\ref{equation:SDPC}). If $\hS$
is rank-one, then $\hS$ is also the optimal solution to (\ref{equation:SDPR}) and one can obtain an
optimal solution $\ts$ to (\ref{equation:QCQP}) by factorizing $\hS = \ts \ts ^T$. Otherwise, if the
rank of $\hS$ is larger than one, then we need to obtain a vector $\ts$ such that the outer product
$\ts \ts ^T$ is close to $\hS$ while $\ts$ remains feasible to (\ref{equation:QCQP}). However, in
general, the obtained feasible solution $\ts$ will not be the optimal solution.

It is natural to use the principal eigenvector of $\hS$, the eigenvector corresponding to the
eigenvalue with largest magnitude, to build the rank-one approximation. Specifically, when
$\rank{\hS} = R$, then $\hS$ has $r$ eigenvalues $\lambda_1 \ge \lambda_2 \ge \cdots \ge \lambda_R >
0$ and eigenvectors $\mathbf{u} _1, \mathbf{u} _2, \ldots, \mathbf{u} _r \in \mathbb{R} ^ N$ such
that the eigen-decomposition is $\hS = \sum _{k = 1} ^R \lambda _k \mathbf{u} _k \mathbf{u} _k ^T =
\mathbf{U} \boldsymbol{\Sigma} \mathbf{U} ^T$, where $\mathbf{U} = \left[ \mathbf{u} _1, \mathbf{u}
_2, \ldots, \mathbf{u} _r \right]$ and $\boldsymbol{\Sigma}$ is the diagonal matrix with
$\diag{\boldsymbol{\Sigma}} = {\left[ \lambda _1, \lambda _2, \ldots, \lambda _r \right]} ^T$. Since
$\lambda _1 \mathbf{u} _1 \mathbf{u} _1 ^T$ is the best rank one approximation of $\hS$ in the
Frobenius norm sense, $\vw = \sqrt{\lambda _1} \mathbf{u} _1$ can be a candidate solution to problem
(\ref{equation:QCQP}), provided that it remains feasible.

Randomization is another way to perform the rank-one approximation. Assume that $\mathbf{v} \in
\mathbb{R} ^R $ is a random vector whose entries are drawn independently and identically according
to the standard Gaussian distribution, i.e., $\mathbf{v} \sim \mathcal{N} \left( \mathbf{0},
\mathbf{I} \right)$, where $\mathbf{I}$ is the identity matrix. Let $\vw = \mathbf{U}
\boldsymbol{\Sigma} ^{1/2} \mathbf{v}$, where $\boldsymbol{\Sigma} ^{1/2}$ is the element-wise
square root of $\boldsymbol{\Sigma}$. A simple calculation indicates that $\expect{\vw \vw ^T} =
\hS$, where $\expect{\cdot}$ returns the element-wise expectation. Furthermore, we have $\expect{\vw
^T \ma \vw} = \expect{\trace{\ma \vw \vw ^T}} = \trace{\ma \hS}$.  Finally, we have $\expect{\vw ^ T
\mb _k \vw} = \trace{\mb _k \hS}$ and $\expect{\vw ^T \mc _k \vw} = \trace{\mc _k \hS}$.  Thus $\vw$
maximizes the expected value of the objective function in (\ref{equation:QCQP}) and satisfies the
corresponding constraints in expectation. In other words, the SDP relaxation in
(\ref{equation:SDPC}) is equivalent to the following stochastic QCQP\@:
\begin{align}
    \hS = \arg \max _{\ms \in \mathbb{S} ^N} \quad 
    & \expect{\vw^T \ma \vw} \notag \\
    \st \quad
    & \expect{\vw^T \mb _k \vw} \le \alpha _k, \quad k \in \mathcal{I}, \notag \\
    & \expect{\vw^T \mc _k \vw} = \beta _k, \quad k \in \mathcal{E}, \notag \\
    & \vw \sim \mathcal{N} \left(\mathbf{0}, \mathbf{S}\right).
    \label{eq:stoc-qcqp}
\end{align}
Such stochastic interpretation of the SDP relaxation provides an alternative way to generate the
rank-one approximated solution to (\ref{equation:SDPC}).

However, both the approximated solutions $\mathbf{w}$ from the eigen-decomposition and the randomized
projection may not be feasible for the original QCQP\@. A feasible solution $\ts$ can be obtained by
projecting the approximated solution $\mathbf{w}$ onto the feasible solution set such that $\ts$ is
the nearest feasible solution to $\mathbf{w}$. For example, the feasible solutions for QCQP classes
I and III are obtained by the element-wise multiplication $\tilde{s} _k = \sign{w _k} \cdot \beta
_k$ ($k = 1, 2, \dots, N$), where $\tilde{s} _k$ and $w _k$ denote the $k^{\mathrm{th}}$ entries of
$\ts$ and $\vw$, respectively~\cite{ye1999approximating, zhang2000quadratic}. As a special case of
QCQP class I, the feasible solutions for QCQPs under binary constraints (i.e., all $\beta _k = 1$)
are obtained via binary quantization $\ts = \sign{\vw}$, where $\sign{\vw}$ returns the signs of all
entries of $\mathbf{w}$~\cite{goemans1995improved}. Alternatively, $\ts = \vw \max _{k \in
\mathcal{I}} \sqrt{ \alpha _k / \left( \vw ^T \mb _k \vw \right) }$ has been used to obtain feasible
solutions for QCQP class II~\cite{Nemirovski1999On-maximization}.

Although leveraging the principal eigenvector of $\hS$ is another simple way of applying
rank-one approximation to $\hS$, such an approach is not suitable for problems featuring
additional constraints. More specifically, when binary constraints are included and the principal
eigenvector is found to meet the inequality constraint, the binary quantization will likely affect the optimality 
and feasibility of the approximate solution. Additionally, as shown in the sequel, the quantized principal
eigenvector provides performance worse than that obtained by the sequence given by our randomized
approach.

\subsection{Approximation Ratio}\label{section:ratio}

The goal of the SDP relaxation (\ref{equation:SDPC}) is to obtain the candidate solution $\ts$ for
problem (\ref{equation:QCQP}) that is as close to the optimal solution $\hs$ as possible. Since any
optimal solution $\hs$ to the QCQP (\ref{equation:QCQP}) can produce a feasible solution $\hs \hs
^T$ to the SDP relaxation (\ref{equation:SDPC}), we have $f ( \hs ) \le f ( \hS)$, where $\hS$
represents the optimal solution to the SDP relaxation (\ref{equation:SDPC}).  Additionally, $f ( \ts
) \le f ( \hs )$ due to the fact that the solution $\ts$ resulting from the relaxation solution
$\hS$ by any method should be feasible to the original problem (\ref{equation:QCQP}). Based on these
relationships, if $\gamma = f(\ts)/f(\hS)$ is the ratio between the objective function for a
feasible solution $\ts$ obtained by a rank-one approximation method and the objective function value
for the SDP relaxation optimal solution $\hS$, then this performance ratio is no smaller than that
for $\hs$ with the same factor:
\begin{align}
\gamma = \frac{f(\ts)}{f(\hS)} \le \frac{f(\ts)}{f(\hs)} \le 1.
\end{align}
The factor $\gamma$ measures not only how good the approximation method is but also how close the
resulting solution is to the optimal solution in terms of the objective function's value.

The SDP relaxation with randomized projection provides guaranteed approximation ratios for many QCQP
problems. For example, such a scheme generates an approximation algorithm for the Maximum Cut
problem (belonging to class I), with $\gamma \ge 0.87$~\cite{goemans1995improved}.

\section{Spectrally Shaped Binary Sequence Design}\label{section:design}

In this section, we develop an efficient method to generate a binary sequence that is based on the
SDP relaxation and randomized projection introduced in Section~\ref{section:QCQP}. A filter
implemented to have such a sequence as its impulse response provides a frequency response with a
bandpass and a notch for the message and interferer bands, respectively. We also provide a
theoretical analysis of the algorithm to show its approximation ratio and the likelihood of
feasibility for the randomized sequences obtained. To improve the performance of the algorithm, we
end the section with a discussion on possible additional criteria to select among the multiple
sequences obtained via the proposed randomization.

\subsection{Design Algorithm}

We desire for the spectrally shaped sequence to provide a passband and notch for the pre-determined
message and interferer bands, respectively. We denote by $\cF _M$ and $\cF _I$ the collection
of all discrete Fourier transform basis elements corresponding to the message band $\Omega _M
\subseteq \{1, 2, \dots, N\}$ and interferer band $\Omega _I \subseteq \{1, 2, \dots, N\}$,
respectively. We also assume that $\Omega _M \cap \Omega _I = \emptyset$, but we do not place any
other restrictions on the message and interferer bands. An optimization-based approach for the design 
of an $N$-point spectrally shaped binary sequence can be written as the QCQP
\begin{align}
\hs = \arg \max _{\vs \in \mathbb{R}^N} \quad 
    f (\vs) &= \left\| \cF _M ^H \vs \right\| _2 ^2 \notag \\
    \st \quad  
    g (\vs) &= \left\| \cF _I ^H \vs \right\| _2 ^2 \leq \alpha, \notag \\
    h_k (\vs) & = s ^2 _k = 1, \quad k = 1, 2, \dots, N,
    \label{equation:BSD}
\end{align}
for some interferer tolerance $\alpha > 0$, where $s _k$ denotes the $k^{\mathrm{th}}$ entry of $\vs$. 
As mentioned in Section~\ref{section:QCQP}, such an integer optimization problem is NP-hard.
Though it is possible to use an exhaustive method that searches over all possible binary sequences
to return the optimal sequence when the sequence length is very small, it is too inefficient and
even impossible to use the exhaustive method when the sequence length is relatively large.

Following the framework prescribed in Section~\ref{section:QCQP}, the SDP relaxation for the QCQP
(\ref{equation:BSD}) can be obtained by noting that $\left\| \cF _M ^H \vs \right\| _2 ^2 = \trace{
\cF _M \cF _M ^H \vs \vs^T}$ and $\left\| \cF _I ^H s \right\| _2 ^2 = \trace{ \cF _I \cF _I ^H \vs
\vs^T}$, providing us with the optimization
\begin{align}
    \hS = \arg \max _{\ms \in \mathbb{S} ^N} \quad 
    f \left( \ms \right) &= \trace{\cF _M \cF _M ^H \ms} \notag \\
    \st \quad 
    g \left( \ms \right) &= \trace{\cF _I \cF _I ^H \ms} \leq \alpha / 2, \notag \\
    h _k \left( \ms \right) &= s _{k, k} = 1, \quad k = 1, 2, \ldots, N,
    \label{equation:SDP}
\end{align}
where $s _{k, k}$ denotes the $k^{\mathrm{th}}$ diagonal entry of $\ms$. Note that we omit the
redundant operations that take the real part of $f \left( \ms \right)$ and $g \left( \ms \right)$
since both $\cF _M \cF _M ^H$ and $\cF _M \cF _I ^H$ are Hermitian and these quadratic functions
will always be real-valued. Note also that the bound on the inequality constraint has been
halved in the SDP relaxation in order to provide a theoretical guarantee later in this section.

Our proposed SDP approximation and randomization for the QCQP is detailed in
Algorithm~\ref{algorithm:design}. After obtaining and decomposing the optimal solution $\hS$ for the
SDP relaxation, a randomly generated vector $\mathbf{v}$ is used to project $\hS$ from a high
dimensional space to a low dimensional space and obtain the approximation vector $\vw _\ell$. A
candidate binary sequence $\ts _\ell$ is then obtained by quantizing the approximation vector $\vw
_\ell$. The algorithm repeats the random projection $L$ times to provide a set of candidate
sequences and finally outputs the sequence that maximizes the message band power while meeting the
requested upper bound for the interferer band power.

\begin{algorithm}[t]
\renewcommand{\algorithmicrequire}{\textbf{Input:}}
\renewcommand{\algorithmicensure}{\textbf{Output:}}
\caption{\sl Binary Sequence Design\/}\label{algorithm:design}
\begin{algorithmic}[1]
    \REQUIRE{message band $\Omega _M$, interferer band $\Omega _I$, interferer tolerance $\alpha$,
    random search size $L$}
    \ENSURE{binary sequence $\hs$}
    \STATE{generate bases $\cF _M$, $\cF _I$ for message and interferer bands}
    \STATE{obtain optimal solution $\hS$ to SDP relaxation (\ref{equation:SDP})}
    \STATE{compute SVD for $\hS = \mathbf{U} \boldsymbol{\Lambda} \mathbf{U} ^T$}
    \FOR{$\ell = 1, 2, \ldots, L$}
    \STATE{generate random vector $\mathbf{v} \sim \mathcal{N} \left( \mathbf{0}, \mathbf{I} \right)$}
    \STATE{obtain approximation by projecting $\vw _\ell = \mathbf{U} \boldsymbol{\Lambda} ^{1/2} \mathbf{v}$}
    \STATE{obtain candidate by quantization $\ts _\ell = \sign{w _\ell}$}
    \ENDFOR{}
    \STATE{select best binary sequence}
\begin{align*}
    \hs = \arg \max _{\ts _\ell : 1 \le \ell \le L} \left\{ f\left(\ts _\ell \right)
    : g\left( \ts _\ell \right) \le \alpha \right\}
\end{align*}
\end{algorithmic}
\end{algorithm}

\subsection{Approximation Performance}

As mentioned in Section~\ref{section:ratio}, the goal to the performance analysis of the spectrally
shaped binary sequence design (i.e., the performance of using candidate sequence $\ts$ as the
approximation of optimal sequence $\hs$) is to evaluate the approximation ratio $\gamma$ such that
any $\ts$ generated in step $7$ of Algorithm~\ref{algorithm:design} satisfies $f \left( \ts \right)
\ge \gamma f ( \hs )$. The larger that approximation ratio $\gamma$ is, the closer that candidate
sequence $\ts$ could be to the optimal sequence $\hs$.

Our binary sequence design has a very similar form as the Class III (cf.\
Section~\ref{section:QCQP}): both contain equality constraints and inequality constraints, and the
characteristic matrices for inequality constraints can be factorized as the multiplication of a
canonical vector and its transpose. Those similarities inspire us to use binary quantization $\ts =
\sign{w}$ after randomized projection in sequence design.

However, the characteristic matrices for the inequality constraints of our sequence design QCQP are 
rarely diagonal, preventing it from belonging to Class III\@. It is impossible for $\cF _I
\cF _I ^H$, the characteristic matrix for the inequality constraint in (\ref{equation:SDP}), to be
diagonal except for the uninteresting case when $\Omega _I = \{1, 2, \dots, N\}$ and $\Omega _M =
\emptyset$, i.e., the interferer band covers the whole spectrum. This causes a discrepancy between
the analysis of our proposed approach and that of Class III QCQPs: when the characteristic matrix
$\mb$ for the inequality constraint is diagonal, the inequality constraint function for the
candidate solution $g ( \ts ) = \ts ^T \mb \ts = \trace{\mb \ts \ts ^T}$ is equal to the inequality
constraint function for the SDP relaxation solution $g ( \hS ) = \trace{\mb \hS}$, since the
diagonal entries of $\ts \ts ^T$ and $\hS$ are the same. In contrast, in our proposed sequence
design algorithm, given that $\cF _I \cF _I ^H$ is not diagonal, we have that $g ( \ts ) = \ts ^T
\cF _I ^H \cF _I \ts$ is not equal to $g ( \hS) = \trace{ \cF _I ^H \cF _I \hS }$, even when the
diagonal entries of $\ts \ts ^T$ and $\hS$ are still the same.

\begin{figure}[t]
    \centering
    \includegraphics[width=0.45\textwidth]{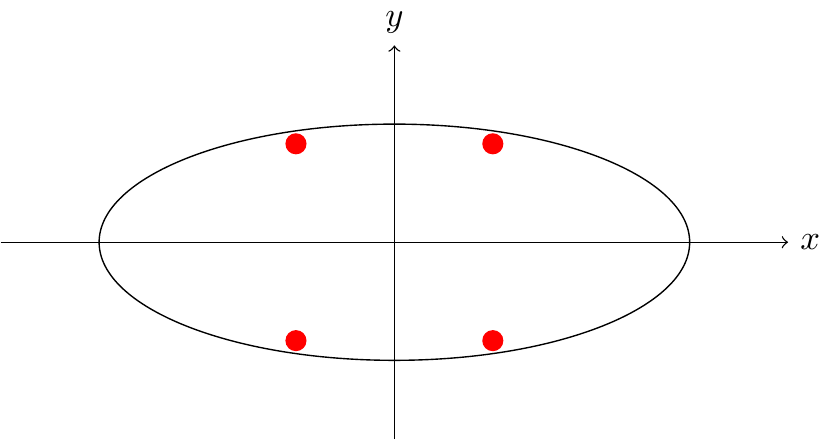}
    \includegraphics[width=0.45\textwidth]{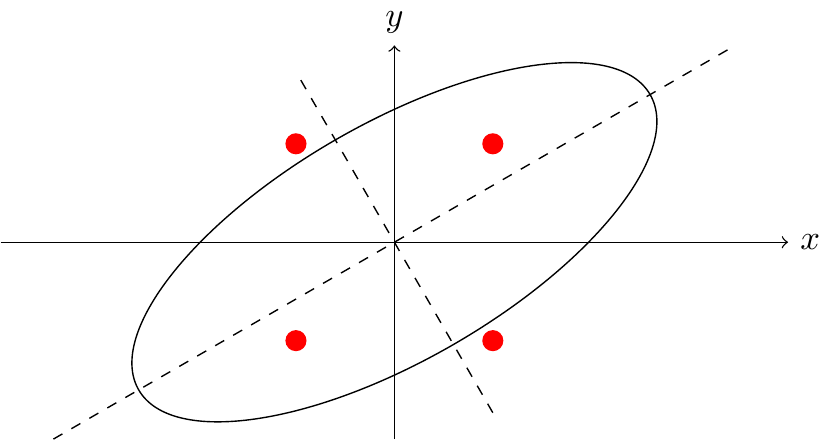}
    \caption{Illustration of feasible sets in (Left) the QCQP Class III and (Right) sequence design.
        Red dots represent the possible binary vectors. Black ellipses represent the bounds of
        inequality constraints.}\label{figure:illustration}
\end{figure}

There is also some geometric intuition behind this difference. Any binary vector obtained via
randomized projection and binary quantization is one of the vertices of a hypercube. To be a
feasible solution, the binary vector must lie inside the set defined by the inequality constraints.
Both $g(\vs)$ in (\ref{equation:QCQP}) and (\ref{equation:BSD}) are quadratic functions and both
characteristic matrices $\mb$ and $\cF _I \cF _I ^H$ are positive semidefinite, so each
inequality constraint defines a set bounded by an ellipsoid in a high dimensional space. The
eigenvectors for $\mb$ and $\cF _I \cF _I ^H$ are the principal axes of the two ellipsoids.
Since $\mb$ is diagonal, the eigenvectors are the canonical vectors and the ellipsoid is symmetric
around each of the axes of the space. If a binary vector lies inside the ellipsoid, then all binary
vectors also lie inside the ellipsoid. In contrast, the eigenvectors for $\cF _I \cF _I ^H$
are rarely canonical, so it is possible for some binary vectors to lie outside the
ellipsoid even when others lie inside. Figure~\ref{figure:illustration} illustrates this difference in an
example two-dimensional space.

In summary, binary sequences $\ts$ resulting from $\hS$ via randomized projection and
binary quantization may not be feasible to the inequality constraint, i.e., $\left\| \cF _I ^H
\ts \right\| _2 ^2 \ge \alpha$. Analyzing the performance of $\ts$ consists of
evaluating the feasibility probability and approximation ratio: the former describes how often
$\ts$ satisfies the inequality constraints and the latter measures how good $\ts$ is
provided that it is feasible.

Intuitively, the feasibility probability of $\ts$ highly depends on $\alpha$ and the rank of
$\cF _I$, which is also the width of the interferer band. As can be seen in
Figure~\ref{figure:illustration}, decreasing $\alpha$ shrinks the ellipsoid defined by the
inequality constraints and therefore fewer binary sequences are contained in the ellipsoid, which
causes a reduced feasibility probability.  Furthermore, a wider interferer band put more strict
constraints on the sequences, which makes it harder for the sequences to to be feasible. These can
be shown in the following theorem, proven in Appendix~\ref{section:proof}.
\begin{theorem}\label{theorem:probability}
Assume that $\hS$ is a solution for the SDP relaxation (\ref{equation:SDP})  and $\ts$ is a binary
vector obtained via randomized projection and binary quantization from $\ms$. Define the ratio
\begin{align}\label{equation:def}
\beta = \frac{\trace{\cF _I \cF _I ^H \arcsin \hS}}{\trace{\cF_I \cF_I ^H \hS}}.
\end{align}
Then, we have
\begin{align}\label{equation:prob}
    \Pb{\| \cF _I ^H \ts \| _2 ^2 \ge \frac{1}{\pi} \left( \beta + 1 \right) \alpha} \le \exp \left(
    - C \frac{\alpha ^2}{K^2} \right).
\end{align}
where $C$ is a constant and $K$ is the column number of $\cF _I$.
\end{theorem}

\begin{figure}[t]
    \centering
    \includegraphics[width=0.45\textwidth]{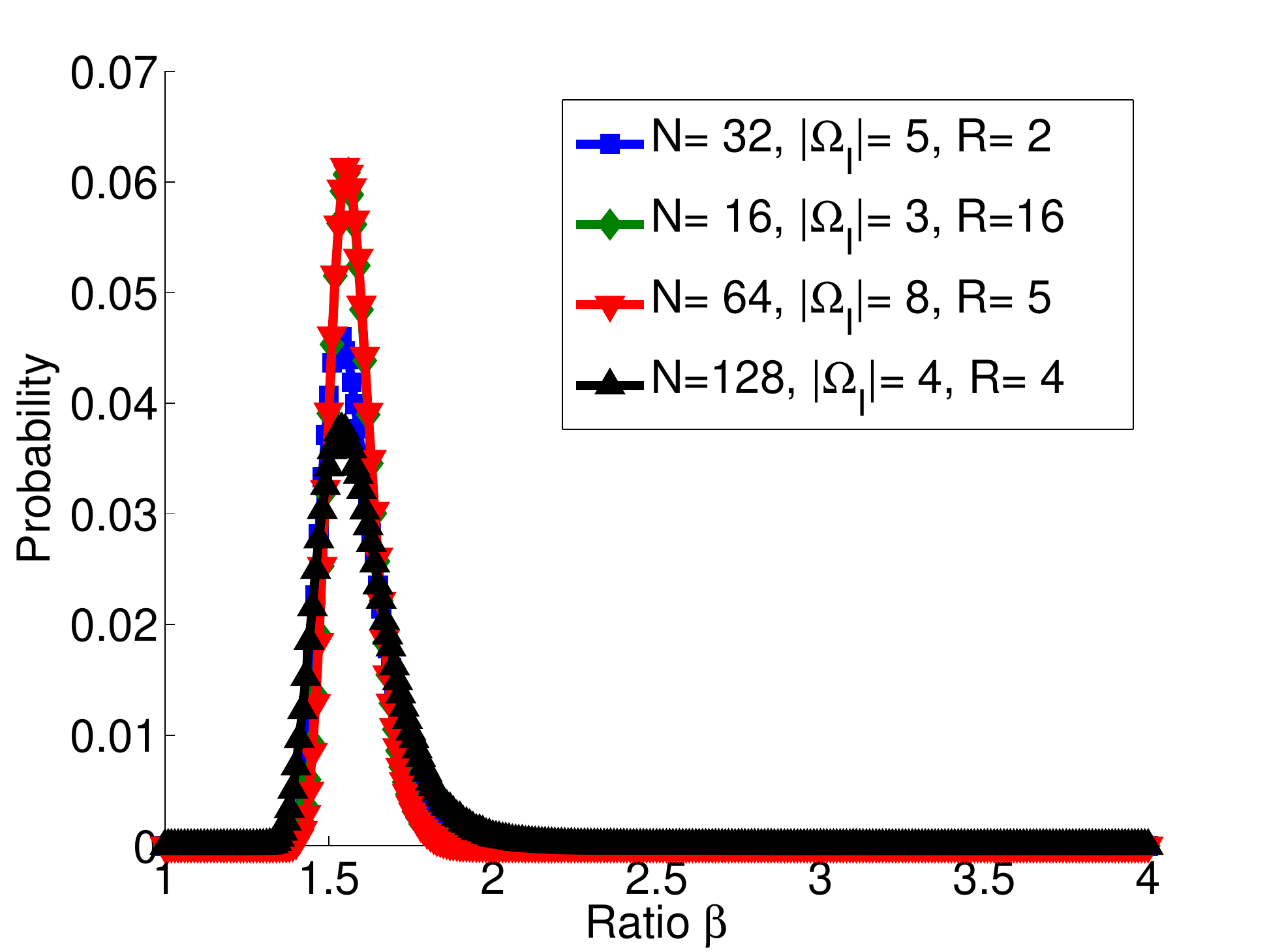}
    \caption{Empirical probability of the ratio $\beta$ between $\trace{\cF _I
    \cF _I ^H \arcsin \hS}$ and $\trace{\cF _I \cF _I ^H \hS}$ in different setting of sequence
length $N$, interferer width $K$, and rank $R$ of $\hS$.}\label{figure:bound}
\end{figure}

It is worthing noting that the ratio $\beta$ depends on the particular solution $\hS$. Furthermore,
it is impossible to obtain a upper bound for $\beta$. To see this, consider
the case when all columns of $\hS$ lie in the null space of $\cF _I$, which would cause
$\trace{\cF _I ^H \cF _I \ms} = 0$. The ratio $\beta$ will be infinite even if $\trace{\cF _I
^H \cF _I \arcsin \hS}$ is very small but not zero. We also evaluate this dependence numerically:
Figure~\ref{figure:bound} shows the empirical probability of the ratio $\beta$ over randomly
generated positive semidefinite matrices $\hS$ for several choices of sequence design problems.
Virtually all instances of the ratio $\beta$ are below $\pi - 1 \approx 2.14$. When this bound on
$\beta$ holds, the result above is reduced to
\begin{align}
\Pb{\| \cF _I ^H \vs \| _2 ^2 \ge \alpha} \le \exp \left( - C \frac{\alpha ^2}{K^2} \right).
\end{align}
Again, note that the reduction of the feasibility bound in (\ref{equation:SDP}) from $\alpha$ to
$\alpha/2$ is necessary to obtain the result above, given the values of $\beta$ that are observed in
practice.

Numerical simulations in the sequel serve as further validation of
Theorem~\ref{theorem:probability}, and confirm the conclusion that the larger that $\alpha$ is, and
the narrower that the interferer band is, the more likely that the sequence $\ts$ will meet the
interferer band power constraint. Additionally, Theorem~\ref{theorem:probability} implies that it is
necessary to generate a sufficiently large number of candidate sequences to meet the feasibility
constraints, as described in Algorithm~\ref{algorithm:design}.

When $\ts$ is feasible to all constraints, it is possible to calculate the approximation
ratio. We claim the approximation ratio by the following conjecture. Such result matches the results
of other QCQPs proved repeatedly in the literature, e.g.,~\cite[Corollary 2.1]{Nesterov1998Global-Quadrati}
and~\cite[Proposition 1]{ye1999approximating}. Though we have found a theoretical proof of the 
following statement elusive, we will verify the conjecture numerically in the sequel.

\begin{conjecture}\label{theorem:ratio}
    Consider a binary sequence $\ts$ obtained via randomized projection and binary
    quantization from $\hS$, which is the solution to (\ref{equation:SDP}). Given that
    $\ts$ meets the inequality constraints, i.e.,  $\| \cF _I \ts \| _2 ^2 \le
    \alpha$, the approximation ratio
\begin{align*}
\gamma = \frac{\| \cF _M \ts \| _2 ^2}{\trace{\cF _M ^H \cF _M \hS}}
\end{align*}
    satisfies $\gamma \ge \pi / 2 - 1$.
\end{conjecture}

Theorem~\ref{theorem:probability} and Conjecture~\ref{theorem:ratio} together guarantee that it is
possible to use the randomized projection and binary quantization to generate feasible binary
sequences with high probability for which the message band power is no less than $\pi / 2 - 1$ of
the optimal power among arbitrary sequences. These two results are the theoretical foundation four
our proposed binary sequence design method.

\subsection{Sequence Selection}

In Algorithm~\ref{algorithm:design}, the final sequence selection step not only excludes candidate
sequences that fail the interferer constraints but also finds a sequence for which the value of the
objective function is as close to the optimal sequence as possible. Intuitively, one would choose
the feasible sequence that maximizes the objective function of (\ref{equation:BSD}), which
corresponds to the sequence with maximal energy in the message band.

\begin{figure}[t]
    \centering
    \includegraphics[width=0.45\textwidth]{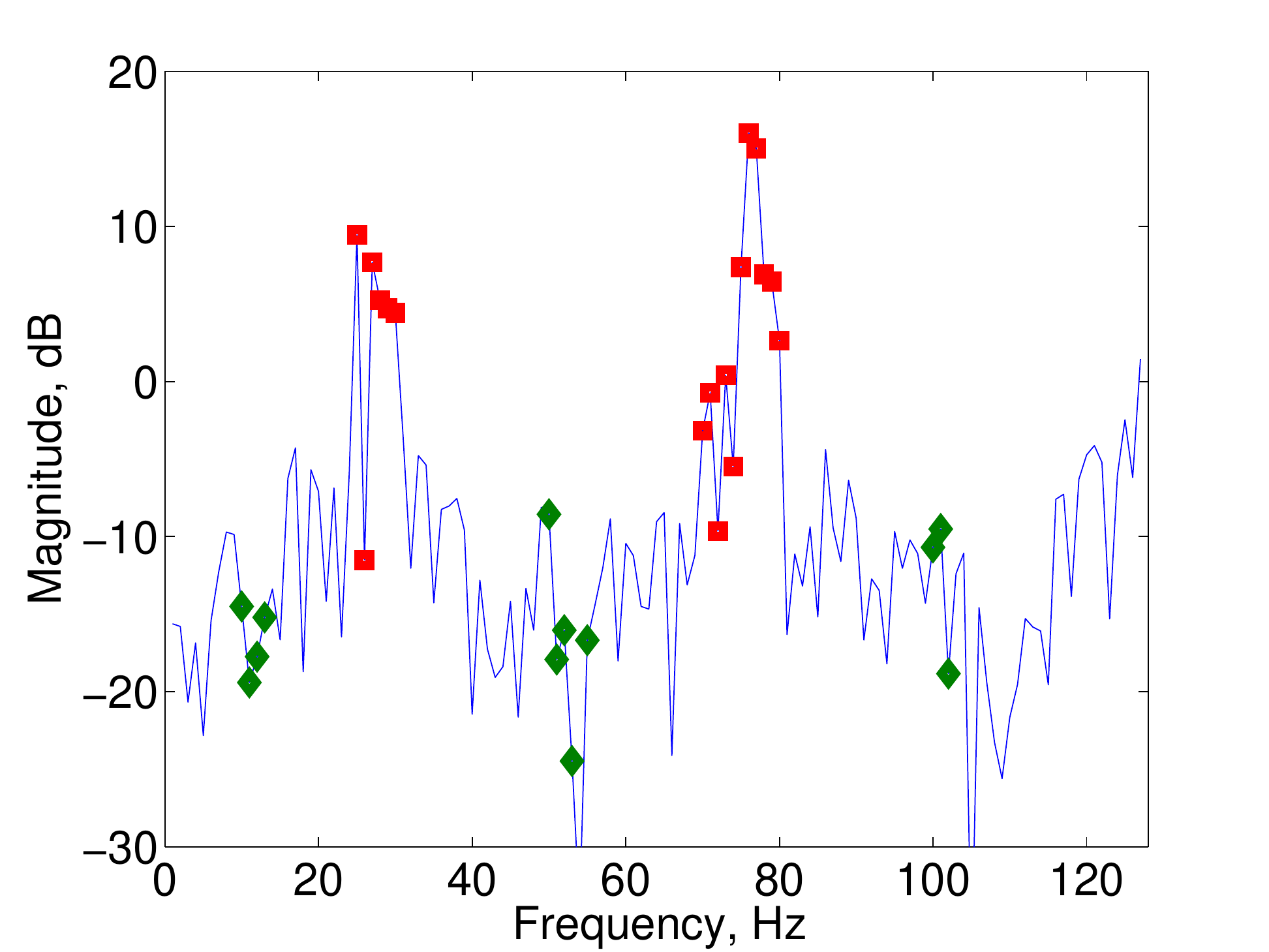}
    \includegraphics[width=0.45\textwidth]{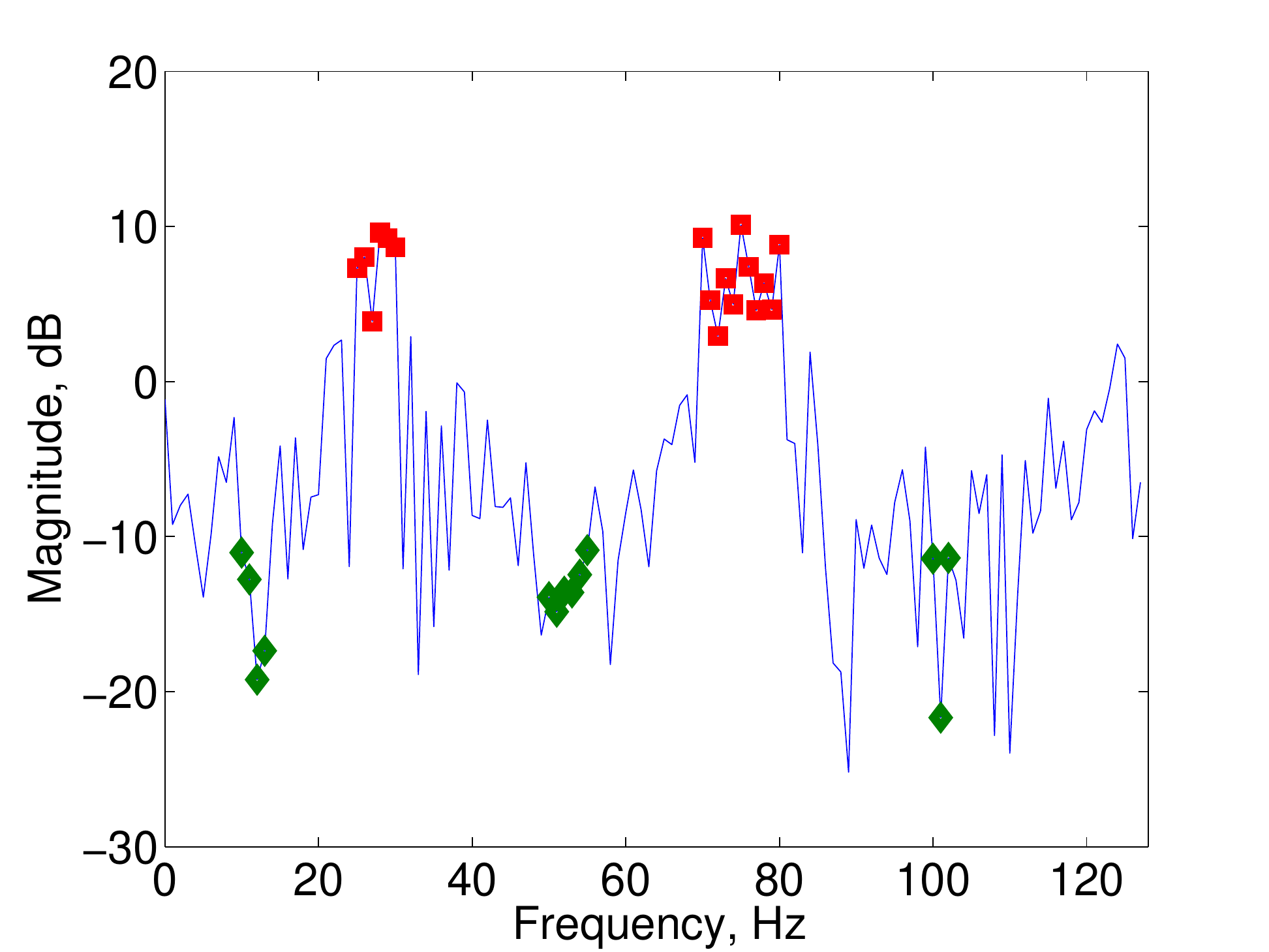}
    \caption{Spectra of example binary sequences that maximize (Left) the message power and (Right)
    the interferer rejection ratio (\ref{equation:MIR}). Green and red markers denote the interferer
and message bands.}\label{figure:score}
\end{figure}

However, the sequence with the largest message power is not necessarily the best suited sequence for
the problem of interest. As shown in Figure~\ref{figure:score}, the sequence selected according to
the message band power maximization often has a large magnitude dynamic range (i.e., the ratio between the
largest magnitude and smallest magnitude), in both the message band and the interferer band.
Additionally, the sequence fails to attenuate the interferer with respect to the message since some
magnitudes in the message band are smaller than some in the interferer band, which potentially
does not allow for successful interference rejection.

To ensure the necessary attenuation, we propose the use of the {\em interferer rejection ratio},
which is defined as the ratio between the minimum magnitude of the spectrum in the message band and
the maximum magnitude in the interferer band, i.e.,
\begin{align}
    \rho \left( \vs \right) := \frac{\min \left| \cF _M ^H \vs \right|}{\max \left| \cF _I ^H \vs \right|},
    \label{equation:MIR}
\end{align}
where the absolute value is taken in an element-wise fashion and the minimum and maximum are
evaluated over the entries of the corresponding vectors. We find that a sequence selection driven by
this criterion provides more amenable spectra for the applications of interest, as shown in
Figure~\ref{figure:score}. Furthermore, we also find in Figure~\ref{figure:score} that the dynamic
range of the spectra in the bands of interest is reduced as well.

\section{Numerical Experiments}\label{section:results}
To test our proposed binary sequence design algorithm, we present two groups of experiments: the
first group provides experimental validation to the two theoretical results from Section~\ref{section:design}; the second group studies the
performance of the obtained sequences in comparison to existing approaches, including the
modifications listed in Section~\ref{section:background} and the exhaustive search when feasible. In
all experiments, the SDP optimization (\ref{equation:BSD}) is implemented using the CVX
package~\cite{Grant2008Graph-implement, Grant2014CVX:-Matlab-Sof}.

\begin{figure}[t]
    \centering
    \includegraphics[width=0.45\textwidth]{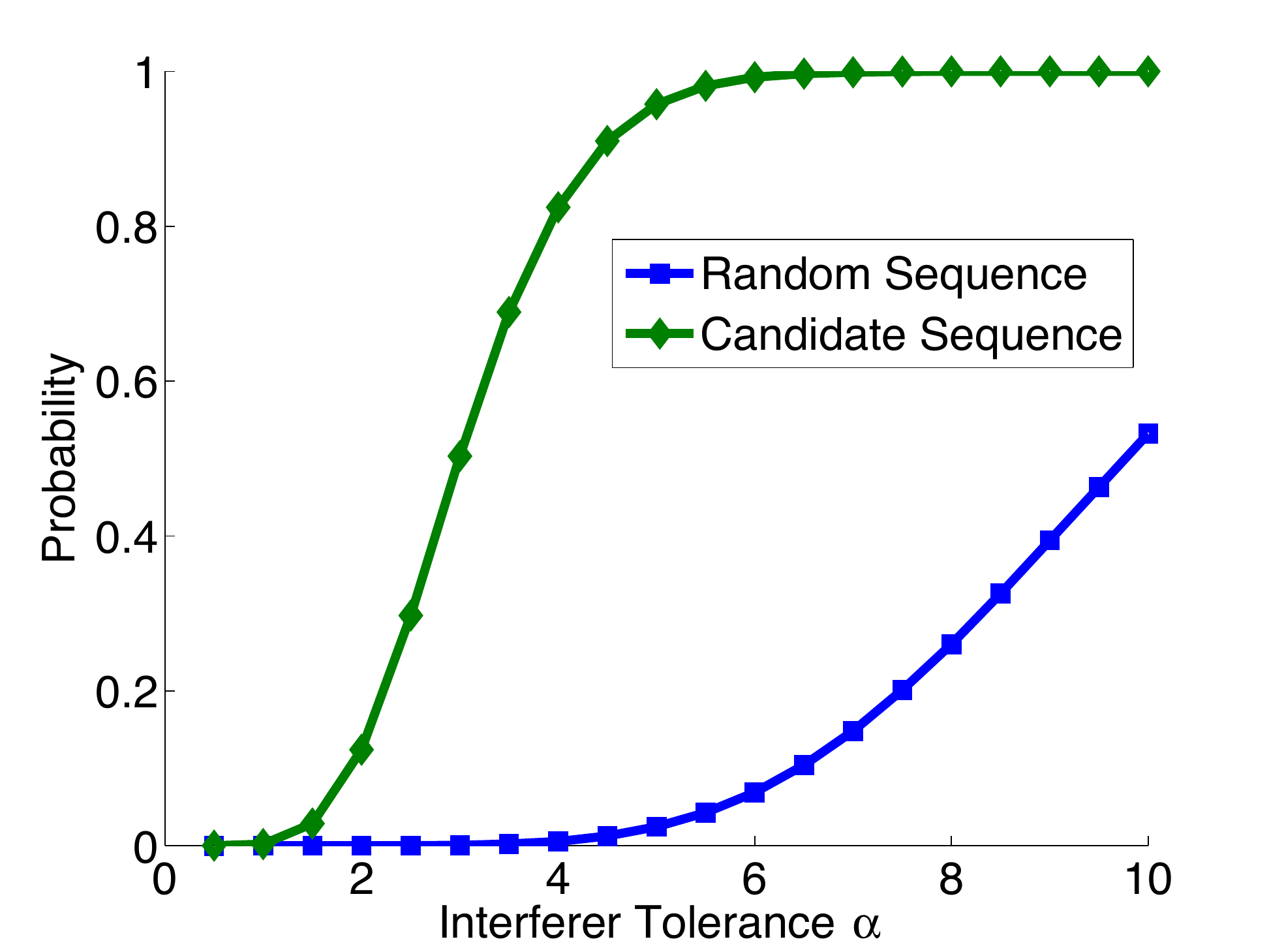}
    \includegraphics[width=0.45\textwidth]{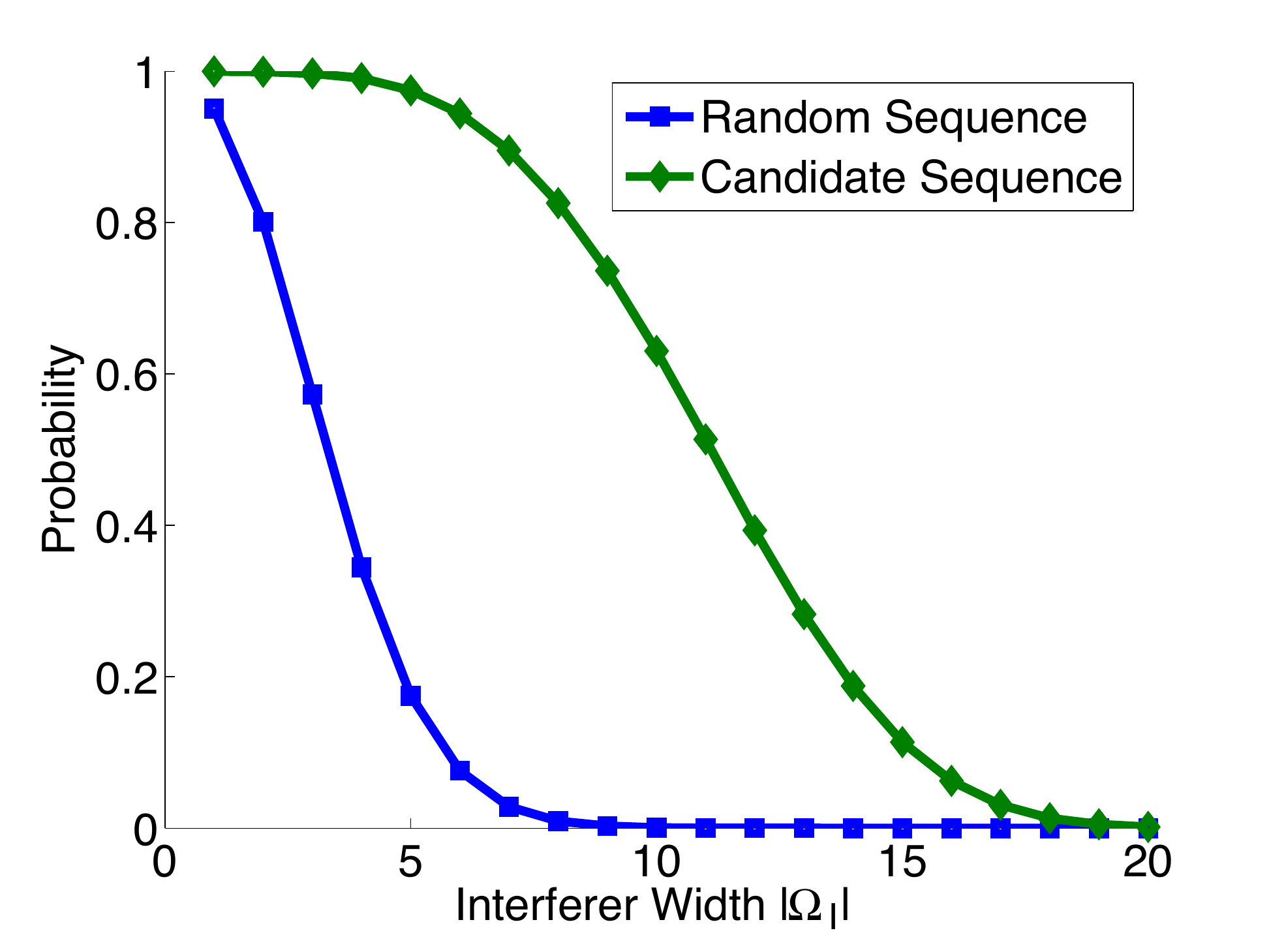}
    \caption{Probability of a candidate sequence satisfying the interferer constraint as a function
    of (Left) interferer tolerance and (Right) interferer bandwidth.}\label{figure:prob}
\end{figure}

In the first experiment, we illustrate the probability that the candidate sequences $\ts$,
obtained according to Algorithm~\ref{algorithm:design}, satisfy the interferer constraint. We set
the sequence length to $N = 128$, and draw $L = 10^6$ candidate
sequences to evaluate the statistical behavior of the algorithm. Figure~\ref{figure:prob} shows the
feasibility probability as a function of the interferer tolerance $\alpha \in [0.5, 10]$ when the
message and interferer bands include the frequencies $\Omega _M = \{25, 26, \dots, 30, 40, 41,
\dots, 45\}$ and $\Omega _I = \{10, 11, \dots, 15, 50, 51, \dots, 55\}$, respectively, and the
feasibility probability, when the message band is $\Omega _M = \{1, 2, \dots, 10, 50, 51, \dots,
60\}$ and the interferer tolerance is $\alpha = 3$, as a function of the interferer width $\left|
\Omega _I \right| \in [1, 20]$ such that the interferer band includes frequencies with indices
$\Omega _I = \{20, 21, \dots, 20+\left| \Omega _I \right|\}$. Both validate the exponential
relationships predicted by Theorem~\ref{theorem:probability}. The blue plots in the figures
correspond to sequences drawn uniformly at random from ${\{-1, 1\}}^N$. The random sequences have
much lower probability to satisfy the interferer constraints than the candidate sequences. This
indicates that it is beneficial to use the combination of an SDP relaxation and randomized
projection to find the feasible sequence.

\begin{figure}[t]
    \centering
    \includegraphics[width=0.45\textwidth]{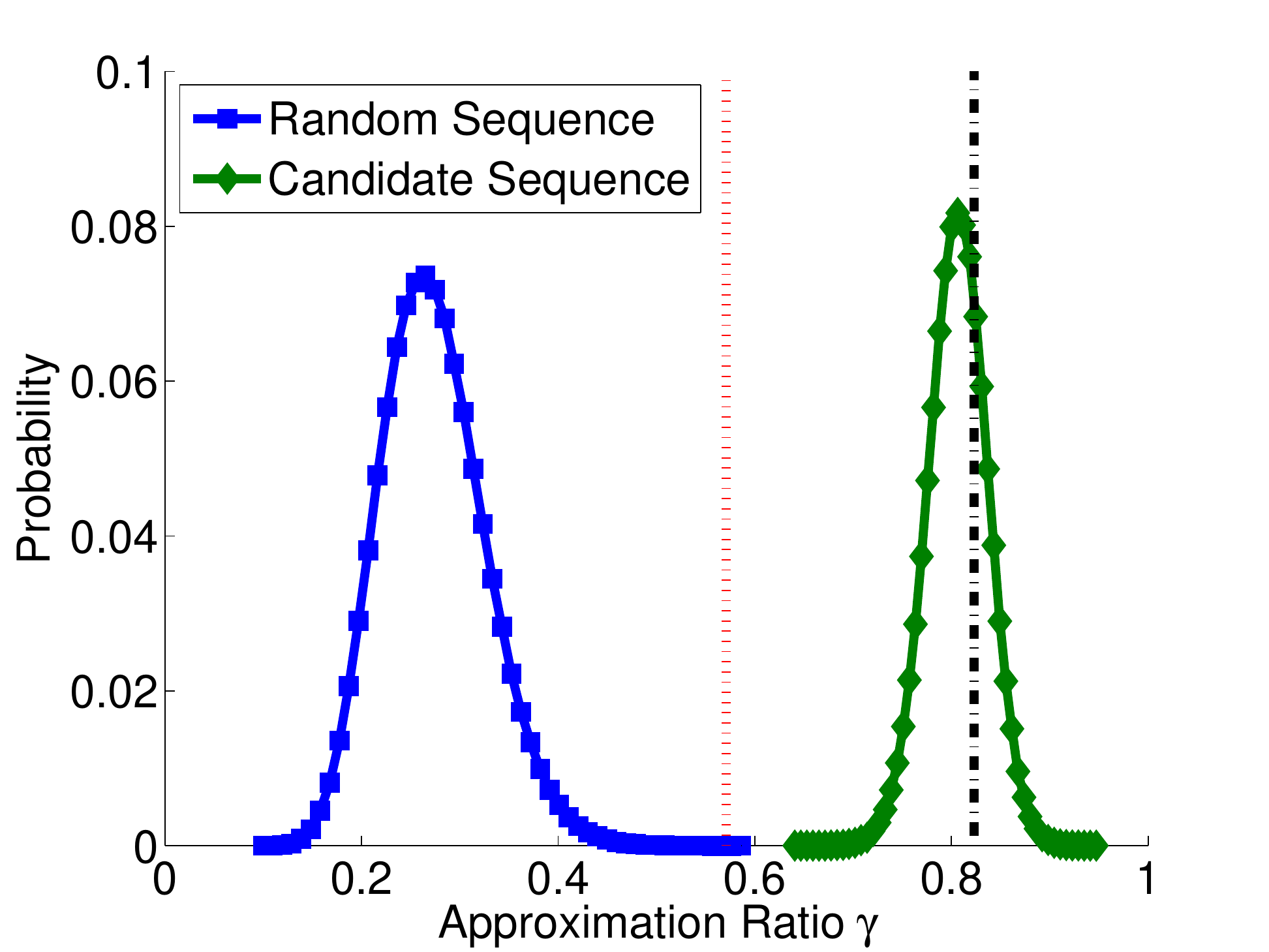}
    \caption{Distribution of approximation ratio $\gamma$ of candidate sequences that are feasible
    to the interferer constraints. The red dotted and black dashed lines represent the bound
predicted by Conjecture~\ref{theorem:ratio} ($\gamma \ge \pi/2 - 1$) and the approximation ratio
corresponding to the quantized principal eigenvector of $\hS$,
respectively.}\label{figure:ratio}
\end{figure}

In the second experiment, we illustrate the distribution of the approximation ratio of the candidate
sequence resulting from the randomized projection and binary quantization. The approximation ratio
corresponds to the ratio of the values of the objective function $f(s)$ from (\ref{equation:BSD})
for the solution $\ts$ obtained from Algorithm~\ref{algorithm:design} to the objective
function $f \left( S \right)$ from (\ref{equation:SDP}) for the solution $\hS$. The setting is
the same as in the previous experiments: $N = 128$,  $\Omega _M = \{25, 26, \dots, 30, 40, 41,
\dots, 45\}$, $\Omega _I = \{10, 11, \dots, 15, 50, 51, \dots, 55\}$, $\alpha = 5$ and $R =
10^6$. We also compare to $R$ random binary sequences with entries drawn from a uniform 
Rademacher distribution. Figure~\ref{figure:ratio} shows that all feasible sequences generated by
Algorithm~\ref{algorithm:design} have approximation ratio $\gamma \ge \pi / 2 - 1$, which is marked 
by the red dotted line; the figure also shows that Algorithm~\ref{algorithm:design} consistently 
outperforms random sequence designs, as expected from the spectral shaping. This numerically 
proves that the sequences obtained from Algorithm~\ref{algorithm:design} meet the approximation 
ratio $\pi / 2 - 1$, as proposed in Conjecture~\ref{theorem:ratio}. Additionally, the results motivate the 
use of random projections rather than only eigendecomposition to obtain the candidate sequences, 
given that some feasible solutions are able to achieve a higher approximation ratio than the 
quantized principal eigenvector of $\hS$, whose approximation ratio is marked by the black 
dashed line in Figure~\ref{figure:ratio}. These numerical results show that the candidate sequences 
obtained from Algorithm~\ref{algorithm:design} have a high probability of satisfying the interferer 
constraints and large message band power, which we can interpret as successful spectrally shaped 
binary sequence design.

\begin{figure}[t]
    \centering
    \includegraphics[width=0.45\textwidth]{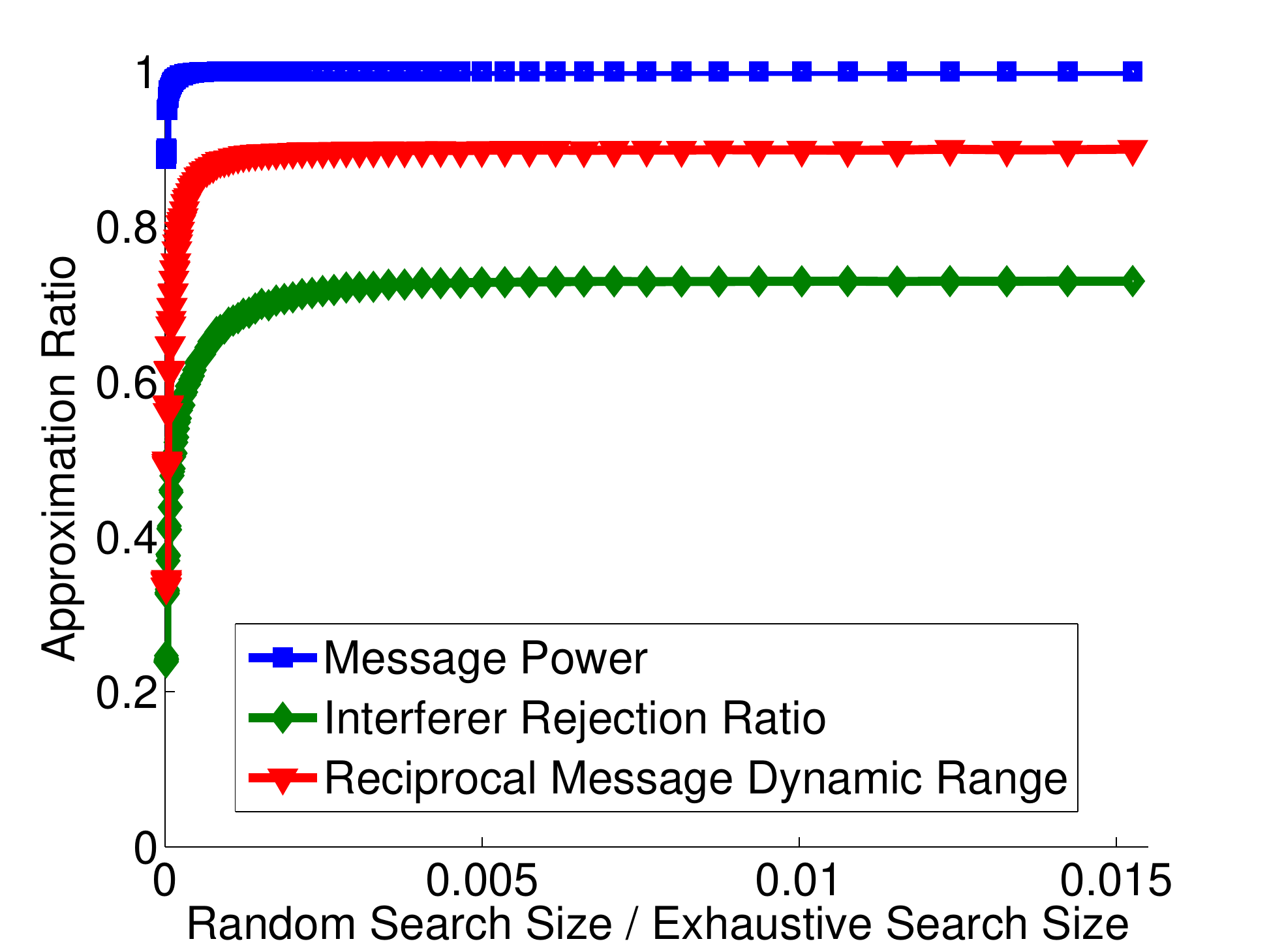}
    \caption{Average ratio of message power and interferer rejection ratio and reciprocal message dynamic range between the sequences
    obtained from Algorithm~\ref{algorithm:design} and the optimal sequences from an exhaustive
search.}\label{figure:compare}
\end{figure}

In the third experiment, we compare the performance of the sequences obtained from
Algorithm~\ref{algorithm:design} versus the optimal sequences obtained by the exhaustive search in
the term of message preservation and interferer rejection. By setting the sequence length $N = 16$, we can feasibly
perform an exhaustive search over all the $2^{16} = 65536$ possible binary sequences. Both the
message and interferer band contain only two frequency bins, and so there are $\binom{8}{2}\times
\binom{6}{2} = 420$ different choices to set the message and interferer band in the spectrum
accordingly, given that message and interferer bands share no common frequency and the spectrum of a
binary sequence is symmetric. Figure~\ref{figure:compare} shows the ratio of the performance metrics
for the sequences obtained by Algorithm~\ref{algorithm:design} over those for the optimal sequences
from an exhaustive search as a function of the size of the random search size $R$ (i.e., the number
of candidate sequences generated in Algorithm~\ref{algorithm:design}) after being normalized by the
size of the exhaustive search space. We use three measures of performance averaged over all $420$
choices: the message band power, the interference rejection ratio~(\ref{equation:MIR}) and the {\em
reciprocal message dynamic range}, which is defined as
\begin{align}
    \chi(s) = \frac{\min \left|\cF _M ^H \vs\right|}{\max \left|\cF _M ^H \vs\right|},
\end{align}
to measure the dynamic range in message band. Noting that each of these metrics can be correspondingly used as the score function in the selection step of Algorithm~\ref{algorithm:design}. The message power of the sequences obtained from Algorithm~\ref{algorithm:design} matches that obtained from exhaustive search, even if $R$, the random search size in
Algorithm~\ref{algorithm:design}, is much smaller than the exhaustive search size. Though the proposed method could not find the sequence with largest interferer rejection ratio, it provides a good approximation with low complexity.

\begin{figure}[t]
    \centering
    \includegraphics[width=0.45\textwidth]{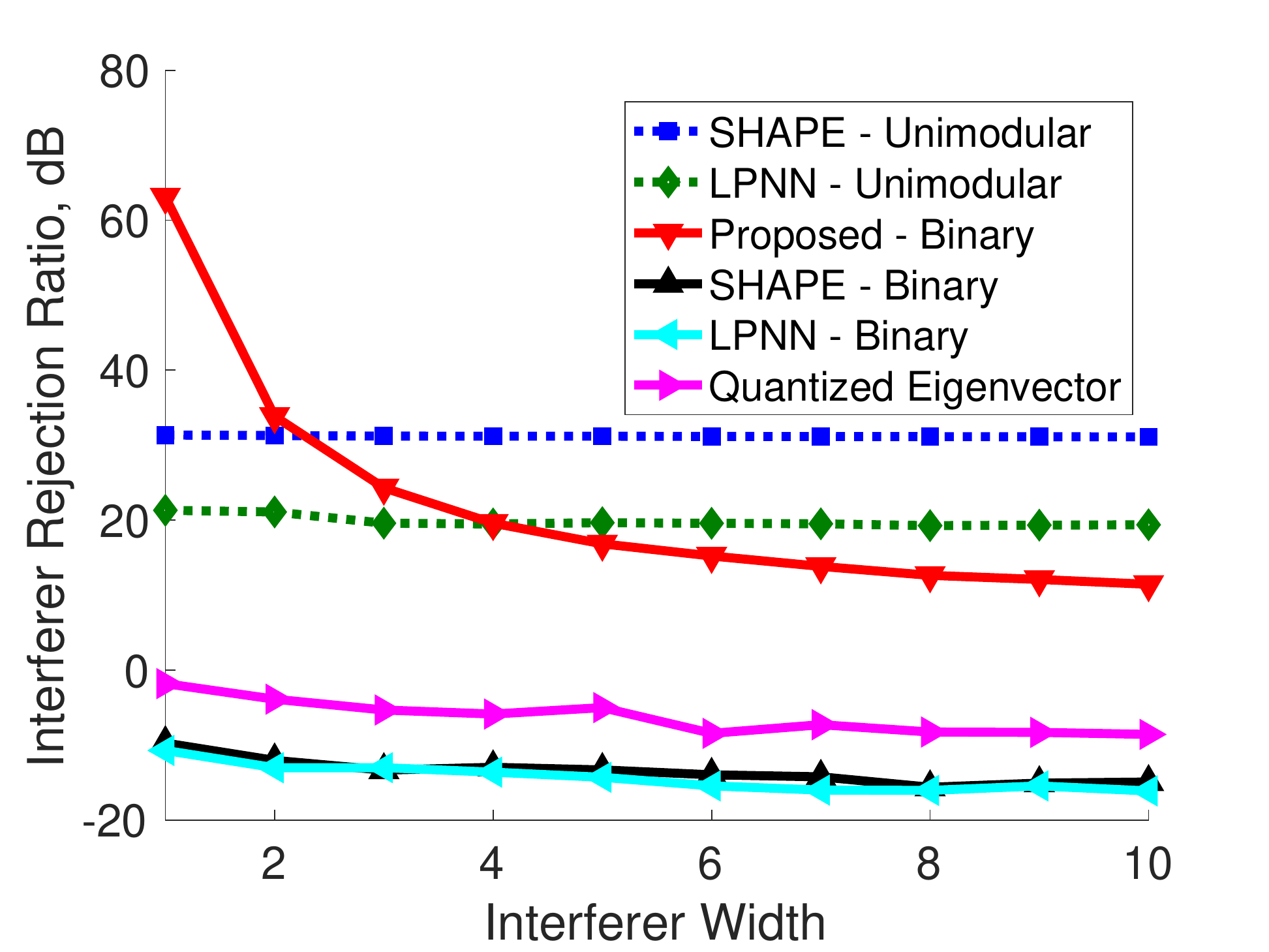}
    \includegraphics[width=0.45\textwidth]{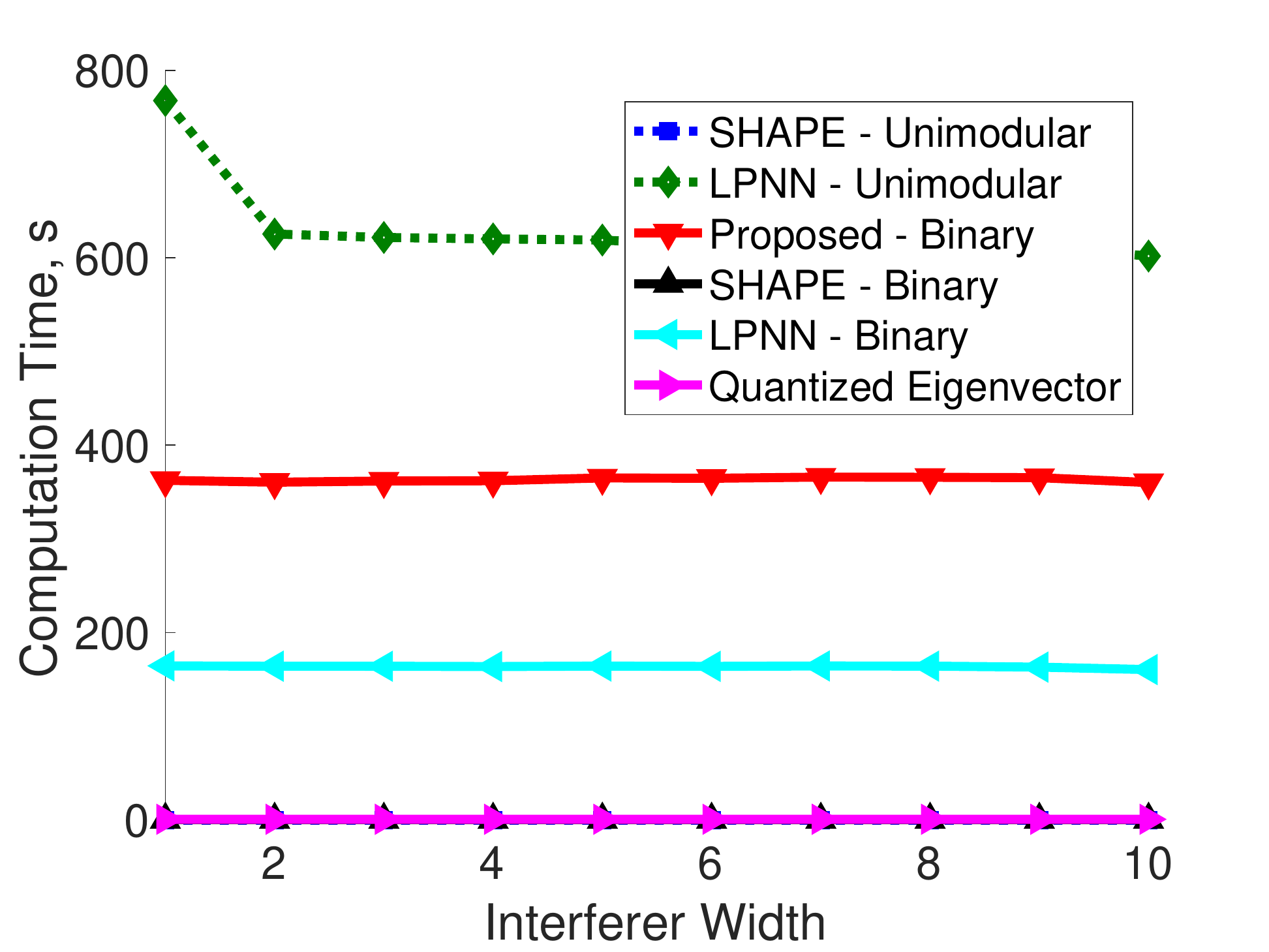}
    \caption{Average interferer rejection ratio (Left) and computation time (Right) of SHAPE and
    LPNN for unimodular sequence and binary sequence and Algorithm~\ref{algorithm:design} for
binary sequence.}\label{figure:performance}
\end{figure}

In the fourth experiment, we compare the performance of the sequences obtained from
Algorithm~\ref{algorithm:design} versus both unimodular and binary sequences from SHAPE and LPNN
algorithms (cf.\ Section~\ref{section:background}) and versus the quantized principal eigenvector approach 
(cf.\ Section~\ref{section:QCQP}) over 100 randomly drawn message and interferer
configurations. The sequence length and the message bandwidth are fixed to be $N= 128$ with $|\Omega
_M| = 10$ and $|\Omega_I|$ varying between 1 and 10. The proposed algorithm chooses the best
sequence from $R = 10^5$ candidate sequences, while the maximum iteration for SHAPE and LPNN
algorithms is $10000$. Figure~\ref{figure:performance} shows the average rejection ratio and
computation time for all tested algorithms. Our proposed algorithm shows the ability to obtain a
binary sequence with a clear distinction in the magnitude of the message and interferer bands. The
performance of our proposed algorithm decreases as the interferer bandwidth becomes larger. Although
it can be expected that the unimodular sequences obtained from both the SHAPE and the LPNN
algorithms provide better interference rejection, surprisingly, our proposed method achieves
performance similar to the unimodular sequences found by existing methods. Furthermore, our proposed
method outperforms their binary-constrained versions, which is indicative of the difficulty of this
more severely constrained problem. Note also that the quantized principal eigenvectors have much worse 
performance than the designed sequences, which is evidence of the benefit provided by the randomized 
projection search included in Algorithm~\ref{algorithm:design}.We finally note that the computation time for 
each algorithm is roughly constant over the interferer widths chosen: our proposed algorithm takes $356$ 
seconds on average while both versions of SHAPE take $0.5$ seconds on average, and the two versions 
of LPNN take $606$ and $181$ seconds on average, respectively. 

\section{Conclusion}\label{section:summary}
In this paper, we propose an algorithm to design a spectrally shaped binary sequence that provides 
a passband and a notch for a pair of pre-determined message and interferer bands, respectively. We
first pose the sequence design problem as a QCQP problem, and combine it with a randomized
projection of the solution of an SDP relaxation (a common convex relaxation) to obtain an
approximation to the optimal sequence in a statistical sense. The candidate sequences obtained by
this method are shown to satisfy the interferer constraints with a probability that depends on the
interferer tolerance and the interferer bandwidth. We numerically show that the candidate sequences
are better approximations (in terms of the objective function value) than sequences obtained by
quantizing the principal eigenvector and than randomly generated binary sequences. Our method 
also outperforms existing approaches for unimodular sequence design that are modified to meet the 
required binary quantization constraint. Our experiments show that for small sequence lengths the 
proposed method is able to obtain the same optimal sequences as the exhaustive search at a 
fraction of the search cost, which shows promise for the use of our randomized method in spectrally 
shaped binary sequence design featuring larger length.

Many questions remain open both on the analysis and possible refinements of our algorithm. For
example, the binary constraint places a significant limitation on the sequence design space. More
flexible quantization schemes that allow for multiple levels in the values of the sequence may
improve the performance of our method while still allowing for a feasible implementation.
Furthermore, one could consider changes to the objective function and the constraints (e.g.,
switching the two) and to the selection score function in order to make the sequences obtained more
relevant to other types of applications. Possible examples include considering the dynamic range of
the message and interferer spectra or the allocation of transmission power to different parts of the
spectrum.

\section*{Acknowledgement}
We thank Dennis Goeckel, Robert Jackson, Joseph Bardin, Wotao Yin, Tamara Sobers and Mohammad Ghadiri Sadrabadi
for helpful comments to the authors during the completion of this research.

\appendix\section{Proof of Theorem~\ref{theorem:probability}}\label{section:proof}
We will use the following results in our proof.
\begin{theorem}[McDiarmid's Inequality~\cite{McDiarmid1989On-the-method-o}]\label{theroem:inequality}
    Let $X = {[X_1, X_2, \dots, X_N]} ^T$ be a family of random variables with $X _i$ taking values
    in a set $\Lambda _i$ for each $i \in \I = \{1, 2, \dots, N\}$. Assume the function $g: \prod
    _{i \in \I} \Lambda _i \rightarrow \R$ satisfies $| g \left(\vx\right) - g \left( \bar{ \vx}
    \right) | \le c_n$ whenever $\vx, \bar{\vx} \in \prod _{i \in \I} \Lambda _i$ differ only in
    their $n ^{\textrm{th}}$ entries for some $n \in \I$. For any $\zeta > 0$, we have
    \begin{align}
        \Pb{g(X) > \expect{g(X)}+\zeta} \le \exp \left( - \frac{2 \zeta ^2}{\sum _{i \in \I} c _i ^2} \right).
    \end{align}
\end{theorem}

To use Theorem~\ref{theroem:inequality} to prove Theorem 1, we need to present some additional results.
\begin{lemma}[{\cite[Lemma 3.2]{goemans1995improved}}]\label{lemma:arccos}
    If $s$ is a binary vector obtained via randomized projection and binary quantization from $S$,
    then for any indices $i, j \in \I$,
\begin{align}
    \Pb{s _i \ne s _j} = \frac{1}{\pi} \arccos \left( \frac{S _{i,j}}{\sqrt{S _{i, i} S _{j, j}}} \right).
\end{align}
\end{lemma}

This lemma provides an important connection between the original binary sequence design and its SDP
relaxation, and allows us to prove the following result, which we have found in the literature without proof.
\begin{lemma}\label{lemma:expect}
If $\vs$ is a binary vector obtained via randomized projection and binary quantization from $\ms$, then
\begin{align}\label{equation:expect}
        \expect{\vs _i \vs _j} = \frac{2}{\pi}\arcsin \ms_{i, j},
    \end{align}
for any indices $i, j \in \I$.
\end{lemma}
\begin{myproof}
Since $\ms$ is a solution for the SDP relaxation (\ref{equation:SDP}), $\ms _{i, i} = 1$ for each $i
    \in \I$, and so $\expect{\vs _i ^2} = 1 = \frac{2}{\pi} \arcsin \ms _{i, i}$. When $i \ne j$,
\begin{align}
    \expect{\vs _i \vs _j} 
                        &= 1 - 2 \Pb{\vs _i \ne \vs _j} = \frac{2}{\pi} \left( \frac{\pi}{2} - \arccos \ms _{i, j} \right) = \frac{2}{\pi} \arcsin \ms _{i, j},
\end{align}
where the second equality is due to Lemma~\ref{lemma:arccos}
\end{myproof}

To use McDiarmid's Inequality, we need to prove the following conditions for binary sequences.
\begin{lemma}\label{theorem:condition}
    If $\vs$ is a binary vector obtained via randomized projection and binary quantization from $S$,
    then $\left| \left\| \cF _I ^H \vs \right\| _2 ^2 - \left\| \cF _I ^H \bar{\vs} \right\| _2 ^2
    \right| \le 4 |\Omega _I|$ whenever $\vs, \bar{\vs} \in {\{-1, 1\}} ^N$ differ only in the $n
    ^{\mathrm{th}}$ entries for any $n \in \I$.
\end{lemma}
\begin{myproof}
    We can express the entries of $\cF _I$ as $a _{k, i} = \frac{1}{\sqrt{N}} e ^{\left(\jmath 2
    \pi (k-1) (i-1) / N \right)}$ ($k \in \Omega _I$, $i \in \I$). Since $\cF _I$ is a
    submatrix of the Fourier orthonormal basis matrix, $\sum _{i \in \I} |a _{k, i}| ^2 = 1$.
    Additionally,
\begin{align}\label{equation:func}
     \left\| \cF _I ^H \vs \right\| _2 ^2 &= \trace{\cF _I \cF _I ^H \vs \vs ^T} = \sum _{i \in \I} \sum _{j \in \I} \sum _{k \in \Omega _I} a _{k, i} ^* a _{k, j} s _i s _j \notag \\
    &= \sum _{i \ne n} \sum _{j \ne n} \sum _{k \in \Omega _I} a _{k, i} ^* a _{k, j} s _i s _j + \sum _{i \ne n} \sum _{k \in \Omega _I} a _{k, i} ^* a _{k, n} s _i s _n + \sum _{j \ne n} \sum _{k \in \Omega _I} a _{k, n} ^* a _{k, j} s _n s _j + \sum _{k \in \Omega _I} a _{k, n} ^* a _{k, n} s _n ^2.
\end{align}
Since $\vs, \bar{\vs} \in {\{-1, 1\}} ^N$ differ only in the $n ^{\text{th}}$ entries, $s _i = \bar{s}
_i$ if $i \ne n$ and $s_n ^2 = \bar{s} _n ^2 = 1$. The first and fourth terms in the right hand side
of (\ref{equation:func}) for $\left\| \cF _I ^H \vs \right\| _2 ^2$ and $\left\| \cF _I ^H \bar{\vs}
\right\| _2 ^2$ are the same. Therefore,
{\small \begin{align}
     \left| \left\| \cF _I ^H \vs \right\| _2 ^2 - \left\| \cF _I ^H \bar{\vs} \right\| _2 ^2 \right| 
    &= \left| \sum _{i \ne n} \sum _{k \in \Omega _I} a _{k, i} ^* a _{k, n} s _i \left( s _n - \bar{s} _n \right) + \sum _{j \ne n} \sum _{k \in \Omega _I} a _{k, n} ^* a _{k, j} s _j \left( s _n - \bar{s} _n \right) \right| \notag \\
    &\le 2 \left| \sum _{i \ne n} \sum _{k \in \Omega _I} a _{k, i}^ * a _{k, n} s _i \right| \left| s _n - \bar{s} _n \right| \le 4 \sqrt{\sum _{k \in \Omega _I} |a _{k, n}| ^2 \sum _{k \in \Omega _I} \left|\sum _{i \ne n} a _{k, i} ^* s _i\right| ^2} \notag \\
    &\le 4 \sqrt{\sum _{k \in \Omega _I} |a _{k, n}| ^2} \sqrt{\sum _{k \in \Omega _I} \sum _{i \ne n} |a _{k, i}| ^2 \sum _{i \ne n} s _i ^2} \le 4 \sqrt{\frac{|\Omega _ I|}{N}} \sqrt{|\Omega _I| \frac{N-1}{N} (N-1)} \notag \\
    &\le 4 |\Omega _I|,
\end{align}}
where the second and third inequalities result from Cauchy-Schwarz inequality.
\end{myproof}

Now, we are ready to prove Theorem 1. From (\ref{equation:expect}), we have 
\begin{align}
    \expect{\left\| \cF _I ^H \vs \right\| _2 ^2}
    &= \expect{\vs ^T \cF _I \cF _I ^H \vs} = \expect{\trace{\cF _I \cF _I ^H \vs \vs^T}} = \trace{\cF _I \cF _I \expect{\vs \vs^T}} = \frac{2}{\pi} \trace{\cF _I ^H \cF _I \arcsin \ms} \notag \\
    & \le \frac{1}{\pi} \beta \alpha, \label{eq:egs}
\end{align}
where the last inequality results from (\ref{equation:def}) and the constraint in
(\ref{equation:SDP}).

By picking $\zeta = \frac{1}{\pi} \alpha > 0$ and applying McDiarmid's inequality for $g(\vs)=\left\|
\cF _I \vs \right\| _2 ^2$ with Lemma~\ref{theorem:condition} and (\ref{eq:egs}), we finally obtain
{\small \begin{align}
    \Pb{\left\| \cF _I \vs \right\| _2 ^2 \ge \frac{1}{\pi} \left( \beta + 1 \right) \alpha} 
    &=\Pb{g(\vs) \ge \expect{g(\vs)} + \frac{1}{\pi} \alpha} \le\exp \left( - \frac{2 {\left( \frac{1}{\pi} \alpha \right)} ^2}{N {\left( 4 |\Omega _I| \right)} ^2} \right) \le \exp \left( - \frac{1}{8 N \pi ^2} \frac{\alpha ^2}{|\Omega _I| ^2} \right). \notag
\end{align}}
This completes the proof of Theorem~\ref{theorem:probability}.

\bibliographystyle{elsarticle-num}
\biboptions{sort&compress}
\bibliography{bibliography}

\end{document}